# ACCELERATION OF STATISTICAL DETECTION OF ZERO-DAY MALWARE IN THE MEMORY DUMP USING CUDA-ENABLED GPU HARDWARE

Igor Korkin Iwan Nesterow
Independent Researchers
Moscow, Russia
{igor.korkin, i.nesterow}@gmail.com

**ABSTRACT**

This paper focuses on the anticipatory enhancement of methods of detecting stealth software. Cyber security detection tools are insufficiently powerful to reveal the most recent cyber-attacks which use malware. In this paper, we will present first an idea of the highest stealth malware, as this is the most complicated scenario for detection because it combines both existing anti-forensic techniques together with their potential improvements. Second, we present new detection methods, which are resilient to this hidden prototype. To help solve this detection challenge, we have analyzed Windows memory content using a new method of Shannon Entropy calculation; methods of digital photogrammetry; the Zipf–Mandelbrot law, as well as by disassembling the memory content and analyzing the output. Finally, we present an idea and architecture of the software tool, which uses CUDA-enabled GPU hardware to speed-up memory forensics. All three ideas are currently a work in progress.

## 1. INTRODUCTION

According to the major antivirus companies, there is presently a significant rise in cyber-attacks, using hidden or rootkit malware (McAfee Labs, 2015a; Wangen, 2015; Symantec, 2015). Three tendencies in malware evolution have become apparent presenting corresponding cyber-security challenges. The first one is the custom-made malware attacks. Applying zero-day or unknown malware makes investigation of cyber security incidents significantly more difficult (Jochheim, 2012). Second, malware uses various anti-forensic techniques, evasion approaches, and rootkit mechanisms, which substantially impair their detection. Finally, investigating this malware has to meet very tight deadlines.

**Well-targeted malware attacks**. Recent cyber security breaches appear to suggest that a wide range of cyber-attacks are well-targeted. Nowadays cyber intrusions are rising at an unprecedented pace. The modern malware such as BlackEnergy malware infiltrated the systems that control critical infrastructure, including oil and gas pipelines, water distribution systems and the power grid. The economic impact of such attacks will be colossal, for example, a cyber-attack on the 50 power plants in the USA could cause $1T in economic damage (Jeff, 2015). US Nuclear Regulatory Commission experienced an 18% increase in computer security incidents in the Nuclear Power Plants. These incidents include unauthorized access; malicious code; and other access attempts (Dingbaum, 2016). These cyber–attacks are already happening. Israel’s Minister of Infrastructure, Energy and Water said that the country’s Public Utility Authority had been targeted by malware. He believes that the terrorist organizations such as Daesh, Hezbollah, Hamas and Al Qaeda have realized this attack (Ragan, 2016). In addition, U.K. government believes that ISIS is planning major cyber-attacks against airlines, hospitals and nuclear power plants (Gilbert, 2015). The recent hackers attack on Kaspersky Lab, which was the first cyber-attack on Antivirus Company and car cyber hijacking look paltry and unimportant. The Stuxnet-like malware’s tendency was reinforced by a cyber-attack on the

Kaspersky Lab (Kaspersky Lab, 2015). In this case, the malware focused on stealing technologies and snooping on ongoing investigations. The CEO said the following: "the cost of developing and maintaining such a malicious framework is colossal. The thinking behind it is a generation ahead of anything we'd seen earlier – it uses a number of tricks that make it really difficult to detect and neutralize. It looks like the people behind Duqu 2.0 were fully confident that it would be impossible to have their clandestine activity exposed" (Kaspersky, 2015). This vulnerability of a respected antivirus company reflects a highly sophisticated level of cyber-attacks. It presents a considerable challenge for zero-day detection for both Windows and Unix-based operating systems (Farrukh, & Muddassar, 2012).

**Anti-forensic techniques**. Malware applies a variety of anti-forensic and rootkit techniques to overcome detection or makes it much more difficult. Currently the abnormal rise of anti-forensic techniques and digital investigators cannot match this challenge (SANS Institute, 2015a). According to Alissa Torres, founder of Sibertor Forensics and former member of the Mandiant Computer Incident Response Team (MCIRT), "Attackers know how forensics investigators work and they are becoming increasingly more sophisticated at using methods that leave few traces behind – we are in an arms race where the key difference is training." (Seals, 2015). This trend has been confirmed by the recent attack on Windows operating system (OS) using malware, which by encrypting itself was able to evade popular debugging tools (The Register, 2015). The authors also underline that "the attack is more likely to bypass security checks" and "the source of the attack is not easily identified by forensics analysis." (Nat, & Mehtre, 2014). According to the McAfee Labs 2016 Threats Predictions "cyber espionage attacks have become stealthier and that they have become more impactful than prior breaches" (McAfee Labs, 2015b). As a result, it is not enough to create a new detection tool. We need to take into account both anti-forensic techniques: current ones and their expected future developments.

**Long-term malware infection**. Malicious activity of stealth malware can result in financial, reputational, process, and other losses. There are several examples of malware, which have been stealing data for years. Spy network Red October (Kaspersky Lab's Global Research & Analysis Team, 2013) collected data from diplomatic, government and science agencies from the whole world for 5 years. Another example was the stealthy, sophisticated Regin malware, which has been infecting computers since 2008; the recent detection of dozens of its modules shows that this spy network is still active (Weil, 2014; Paganini, 2015). Such a long-term malware infection is completely unacceptable for all business.

Modern antiviruses do not cover these three aspects of new malware and have no ability to react swiftly against such highly sophisticated malware. Thus actual cyber security threats demand a complex review of all existing hidden malware detection methods.

**Aim of this project**. There is a need to develop new methods of detecting hidden malware, which will be resilient to existing anti-forensic techniques such as rootkit countermeasures.

In this paper, we present a research project which seeks to detect zero-day malware in the memory dump under deliberate countermeasures. By applying the synthesis of new methods, we can detect unknown malware in the memory at a very early stage and so prevent their negative sequelae.

**Motivation**. This paper was inspired by the book 'The Art of Memory Forensics' (Ligh, Case, Levy, & Walters, 2014) and the preliminary version of the book 'Rootkits and Bootkits: Reversing Modern Malware and Next Generation Threats' (Matrosov, Rodionov, & Bratus, 2016) and other papers. We were also inspired by helpful comments on the rootkit detection system MASHKA (Korkin, & Nesterov, 2014) from Luka Milkovic, Nicolas Ruff, Giovanni Vigna, Stefan Vömel and Ibrahim (Abe) Baggili.

This paper consists of 6 sections.

Section 2 is devoted to the comparative analysis of the methods used to detect stealth software. It describes how these methods work and what their vulnerabilities are. The combination of existing anti-forensic techniques shows that hidden malware can overcome all popular stealth software detection methods.

Section 3 contains the several scenarios of further improvements of anti-forensic methods. The idea of the most hidden driver or highest stealth malware (HighSteM), which overcomes all popular detection methods and their possible development. HighSteM functionality will be given; HighSteM can acquire sensitive information using memory access without any interaction with OS functions. The example of the keyboard keylogger, which works in this way, is discussed. As a result, we can simulate the most difficult case scenario for detection, and this will be used in further research of its detection.

In section 4 HighSteM detection methods are presented. We will present a variety of ideas to solve the detection challenge, using methods of digital photogrammetry, the Zipf Mandelbrot law, and methods of artificial immune systems as well as by disassembling the memory content and analyzing the output. These methods will be analyzed in terms of their vulnerability to possible anti forensic techniques. Developing, testing and applying these new methods require a huge amount of computational resources, which are not accessible to the vast majority of laboratories. To solve this time consuming task we proposed using modern graphics cards or CUDA-enabled GPU hardware.

Section 5 explores the idea of signature search optimization. As a basis for signature search the algorithm from the rootkit detection system MASHKA was chosen. The suggestion is to combine the facilities of CPU and GPU so that the part with calculating and linear search will use the number of cores of GPU. This helps to significantly accelerate linear search in the memory dump.

Section 6 contains the main conclusions and further research directions.

## 2. ANALYSIS OF THE DRIVER DETECTION APPROACHES IN THE MEMORY DUMP

This project focuses on the detection of hidden drivers, as they have many opportunities to conceal themselves and because of operating in a high privilege mode they can affect the OS and antivirus software. Hidden drivers features are commonly used in spyware (Paganini, 2015; Musavi, & Kharrazi, 2014) and so the priority task for cyber-security is the detection of hidden drivers in the Windows and Unix-based OS. Further analysis will be carried out for the most popular Windows OS, but all results can be adopted for Unix-based OS as well.

This analysis of the more popular methods for detecting hidden software and their resilience to anti forensic techniques should be of great interest to cyber-security experts. This analysis of detection methods and corresponding countermeasures will be based only on publicly available information sources: books, papers in scientific journals, conference presentations, blog posts as well as discussions in forums.

The explanation will be provided through the increasing number of drivers' manipulations for self-concealment. Each detection method will include the corresponding rootkit techniques to hinder the malware or prevent its work. In addition, every following analyzed method will be resilient towards countermeasures of the previously analyzed method.

### 2.1. Classification of Methods Used to Reveal Drivers

The creation of resilient detection methods requires answers to the following questions:

- How drivers work in Windows OS and how they can be hidden;
- How to detect hidden drivers;
- How to predict and face anti-forensic techniques.

To formulate the prerequisites of driver detection the following topics will be covered in the next section:

- The virtual memory content after a driver has been loaded;
- The details and vulnerabilities of a Windows built-in tool, which collects data about installed drivers;
- Analysis of the resilience of alternative approaches to detect drivers.

In addition, the classification of the different methods used to detect drivers will be given.

According to Blunden, (2012), before a kernel mode driver has been started, its executable file is loaded in the memory, the driver's information is added in to several OS' linked lists and it is after that, the DriverEntry function is executed.

Consider the situation with three loaded drivers A, B, and C. Figure 1 shows the content of the virtual memory prior to the loading of these drivers. To simplify the model, on the top of the figure, there are only two OS' linked lists with three drivers' structures, which correspond to the three loaded drivers. At the bottom there are three executable driver files. All these objects may be used as fingerprints to detect drivers.

According to Russinovich, Solomon, & Ionescu, (2012) the Windows built-in tool uses one of these lists to receive information about loaded drivers. Assuming that list #1 is the list utilized by the built-in tool, e.g. called NtQuerySystemInformation with SystemModuleInformation (11) information class. However as this is a list-based mechanism it is vulnerable to anti-forensic techniques. According to Blunden, (2012), by unlinking a driver's structure from this list it is possible to conceal the driver from the built-in tool. This unlinking attack is also known as Direct Kernel Object Manipulation (DKOM) and this technique yields two positive results. The first is that the built-in tool is not able to find this driver; and second is that Windows OS and the hidden driver will continue to work correctly.

Thus the hidden driver is one of the main cyber security threats, and to eliminate it a reliable method of detection is required to detect all hidden drivers. After all the stealth drivers are detected the incident response can be carried out, using special pre-filing investigations such as reverse-engineering (Matrosov, Rodionov, & Bratus, 2016), although a review of this topic is beyond the scope of this paper.

The most common technique to discover a hidden driver or any rootkit is to apply the cross-view approach, which checks equality between two drivers' lists. The first list was created by the built-in tool and the second list uses one of the alternative drivers' detection approaches.

Alternative detection approaches can be classified into two categories according to the subject of the search: first by using OS driver structures and second by using the content of the driver's files in the memory. Their classification is given in Figure 2. The first approach can be further sub-divided into two groups: based on the links between structures and a signature based search of drivers structures.

The second approach can also be further partitioned into two subsets: signatures and a statistical based approach to detect the driver's files in the memory. Each of these methods will be analyzed. The new detection approach l proposed at the end of this paper, is based on the statistical-based approach.

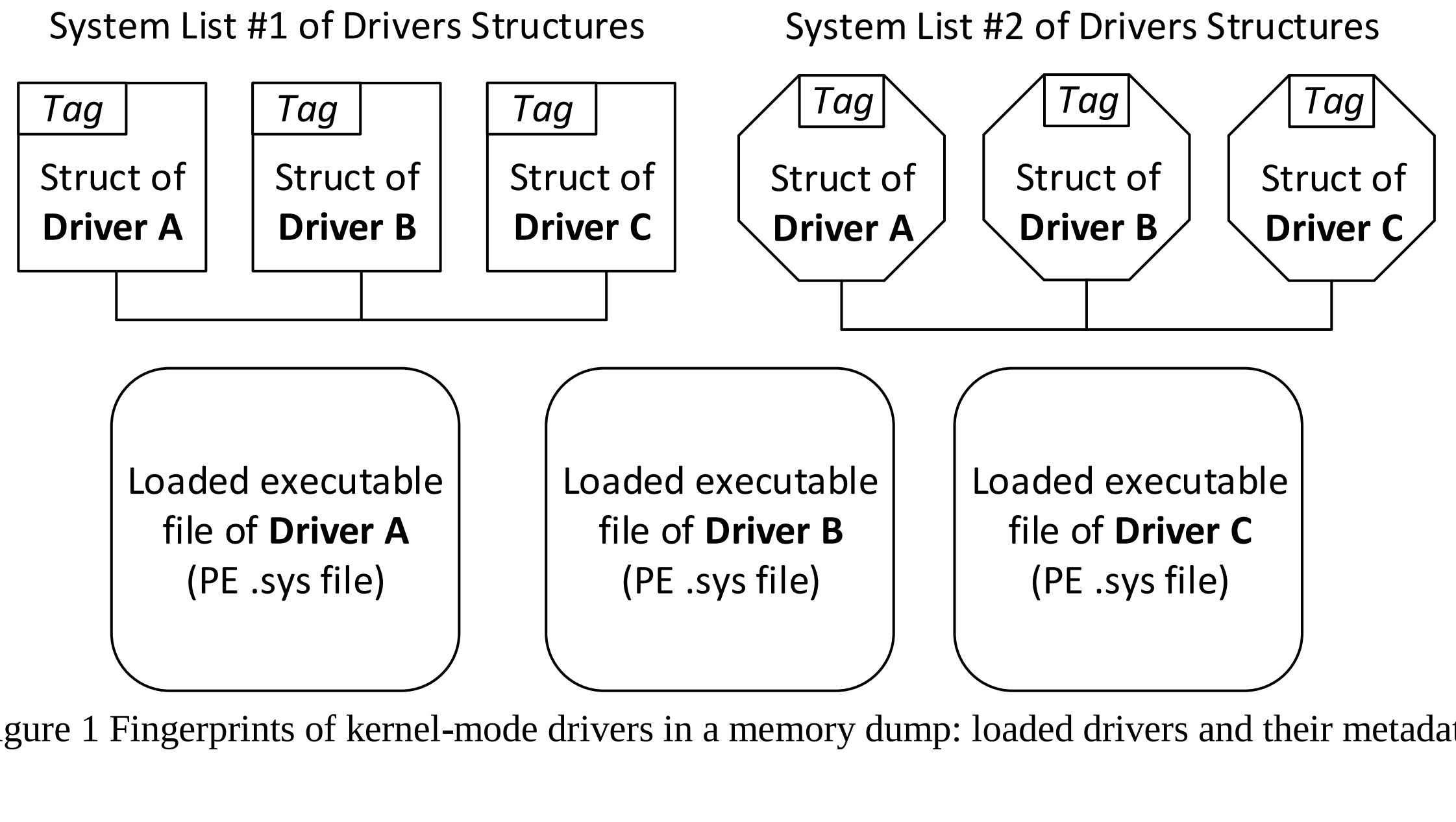

Figure 1 Fingerprints of kernel-mode drivers in a memory dump: loaded drivers and their metadata

### 2.2. Approaches Based on the Links between Structures

These methods receive the list of drivers by walking through the various linked lists of driver structures in the memory.

We will be focusing on the most popular OS's linked lists, which are used to detect drivers (Vomel, & Lenz, 2013):

- List of drivers modules also called PsLoadedModuleList;
- List of kernel mode threads from 'System' process;
- Drivers objects list from the object directory also called ObjectDirectory;
- List of recently unloaded drivers;
- Service record lists in the memory also called ServiceDatabase;
- Database of installed services in the system registry.

We will shortly discuss each of these lists and techniques used to hide data from each of them.

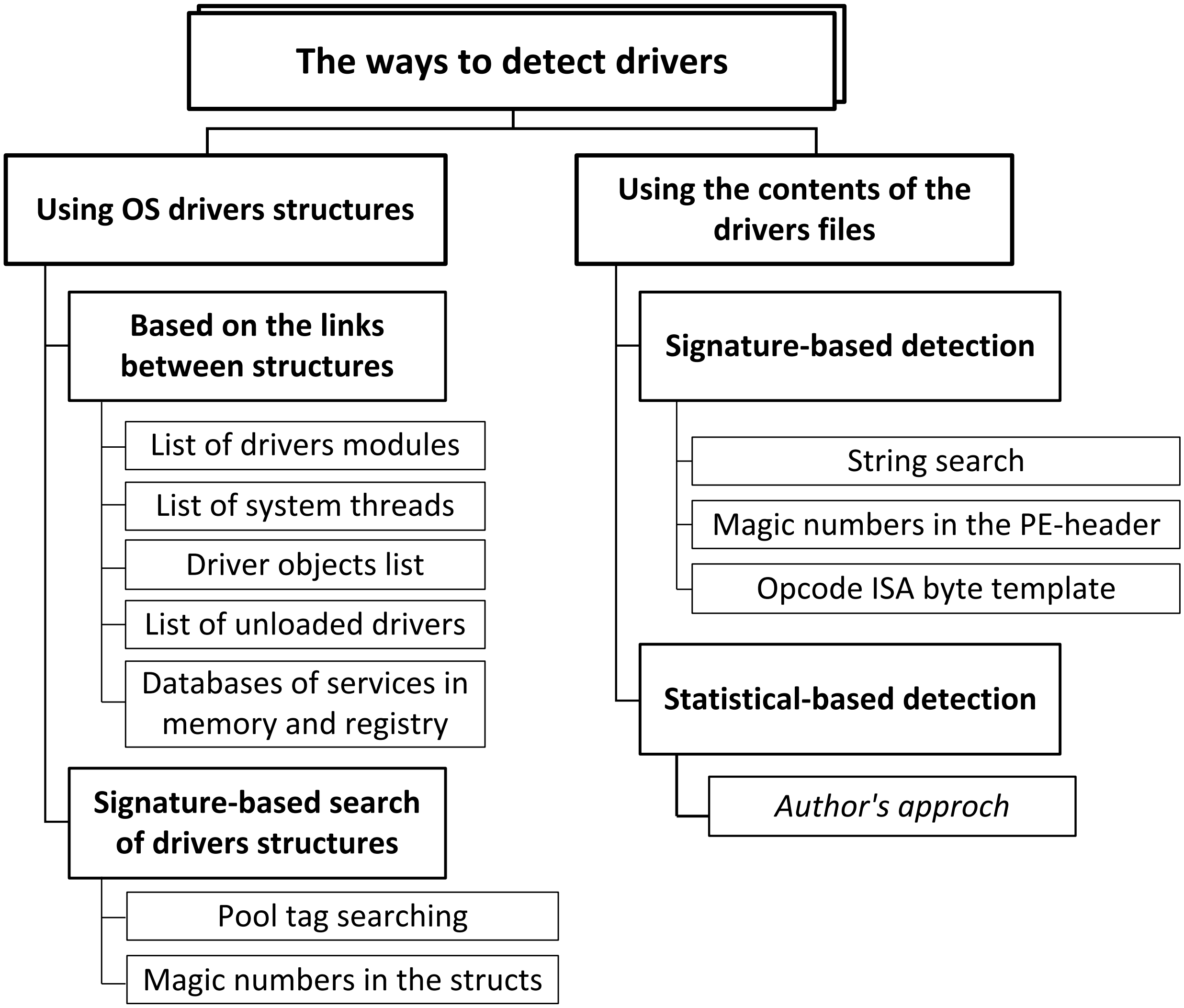

Figure 2 Classification of methods to detect drivers

**PsLoadedModuleList**. The first and the most famous list PsLoadedModuleList, which is used by ZwQuerySystemInformation function or, in other words, by the Windows built-in tool. This is double linked to the list of KLDR_DATA_TABLE_ENTRY structures, which contain the data about each kernel module. The Volatility's *modules* plugin uses this list. It is possible to hide a driver structure from this list by using DKOM unlinking. The details of this anti-forensic technique are given in Hoglund, & Butler, (2005) and also in Tsaur, & Wu, (2014a) and realized in the DhpHideModule function of Dementia, the proof of the concept memory anti-forensic toolkit by Milkovic (2012).

**Kernel-mode threads**. According to the book (Ligh, Case, Levy, & Walters, 2014a) the process 'System' with PID 4 includes the list of kernel mode threads. This list contains the ETHREAD structures. This idea is also used in the Volatility's *orphan* plugin. However, an intruder is able to prevent detection using this list by rewriting sensitive information from the corresponding ETHREAD structure, e.g. field StartAddress. As a result, the modified structure will be of no use for detection. This anti-forensic technique will be referred to as DKOM patching.

DKOM patching is a widely used technique. Here is an idea of how to hide EPROCESS structure – "instead of removing the process from the linked list, *replace its content* and technically the process should be "hidden", means it will still show up on taskmanager etc but when you try to close or dump it, it will close/dump the dummy process instead". (Ch40zz, 2015)

**ObjectDirectory**. The third list is the Windows Object Directory list or the object tree, which contains structures of different objects. In this paper, we are focusing on DRIVER_OBJECT structures. The details of how this tree is organized is given in Zhanglinfu2000 (2013) and Probert (2004). The documented function NtQueryDirectoryObject uses this tree to get the information about the specified directory object (MSDN, n.d.-a). However, the driver is able to conceal the corresponding DRIVER_OBJECT structures by DKOM unlinking. The details of this anti forensic technique are given in (Tsaur, & Yeh, 2015; Jakobsson, & Ramzan, 2008; Tsaur, & Wu, 2014a). The attack of hiding a driver from an object directory is implemented by the StealthInitializeLateMore function in the AntiMida by Pistelli (n.d) and in the Zion open source rootkit by Chen (2008).

**Recently unloaded drivers**. This list includes the names, times, the start and finish addresses of recently unloaded drivers (Ligh, Case, Levy, & Walters, 2014b). This list is mentioned in the following papers (MJ0011, 2009; SANS Institute, 2015b) as well as in the forum discussion (KernelMode.info, 2012). The Volatility's *unloadedmodules* plugin uses this list. According to the Windows Research Kernel (WRK) the variable PUNLOADED_DRIVERS MmUnloadedDrivers stores the address of the top of this list, and by using MmLocateUnloadedDriver function it can locate the specified virtual address in the list (Microsoft, n.d.-a; Microsoft, n.d.-b). We can also get this address by calling ReadDebuggerData function with parameter DEBUG_DATA_MmUnloadedDriversAddr (MSDN, n.d.-b). However, an attacker can apply DKOM patching and overwrite sensitive fields in the corresponding structure, and therefore it makes the list unusable for detection.

**ServiceDatabase in the memory**. According to (Ligh et al, 2014c) after the installation of a driver using Service Control Manager (SCM) corresponding information is added to the two lists. The first is the double linked list of service record structures in services.exe also known as ServiceDatabase. The second list is located in the system registry. Thus it is possible to get information about loaded drivers from ServiceDatabase using the EnumServicesStatus function with parameter SERVICE_KERNEL_DRIVER (MSDN, n.d.-c). The detailed analysis of SCM mechanisms is explained is several recent books (Stuttard, Pinto, Ligh, Adair, Hartstein, & Richard, 2014; Ligh, Case, Levy, & Walters, 2014d) as well as in the blog of D.Clark (Clark, 2014). The ServiceDatabase details and the many ways to conceal the service record structure and this is in the context of Blazgel Trojan in the book (Ligh, Adair, Hartstein, & Richard, 2010), which hides services by DKOM unlinking. Further examples of how to unlink a structure of hidden service are given in Wineblat, (2009); Mask, (2011); Louboutin, (2010).

**Database of Installed Services in the system registry**. As mentioned before if a driver is loaded using SCM, this information will also be duplicated in the system registry. The list of registered kernel mode drivers and user mode services is located in the following registry path "HKEY_LOCAL_MACHINE\SYSTEM\CurrentControlSet\services"(MSDN, n.d.-d; Microsoft, n.d.-c). The detailed description of this list is here (OSR Online, 1997). We can get Information from this list, it can also be obtained by using registry API function RegEnumKeyEx. However this list is also susceptible to intruder countermeasures such as deleting the corresponding registry key, which contains the information about a hidden driver. An example of removing the driver-related information in the registry is given in Tsaur, & Wu, (2014b; ZCM Services, 2010).

Table 1 summarizes the various analyzed approaches to detect drivers and their vulnerabilities. The first column gives the list and the respective structure names. The second column contains the software tools, which apply these lists. The last column shows the anti-forensic techniques, which can defeat or overcome the detection.

As a result, we can conclude that all approaches based on the links between structures are susceptible to anti-forensic techniques such as DKOM unlinking, DKOM patching and removing registry keys. Next, the paper will review the driver detection methods, which are resilient to these countermeasures, although they may have other weaknesses.

### 2.3. Signature-based Search of Drivers Structures

When a driver structure is concealed by using DKOM patching and all informative fields are rewritten then this structure is useless for detection. However, if a structure is hidden only by DKOM unlinking it can be revealed by a signature based search.

Signature-based search of driver structures is based on the fact, that driver structures have typical fragments or in other words signatures, which are the same for all structures from one list. We can reveal all drivers structures by using byte-to-byte signature search regardless of whether or not a structure is unlinked. Signature based search can retrieve data and does not rely on API calls that can be subverted. Signature based search in some cases can be sped up significantly using the fact, that driver structures, e.g. DRIVER_OBJECT, are located closely to each other in the memory (Korkin, & Nesterov, 2014).

Table 1 Summary table of the detection approaches based on the links between structures and their vulnerabilities

<table>
<tr><th>List and structure names</th><th>Examples of lists using</th><th>Examples of Attacks</th></tr>
<tr><td>PsLoadedModuleList,<br>KLDR_DATA_TABLE_ENTRY</td><td>• ZwQuerySystemInformation<br>• Volatility's modules plugin</td><td rowspan="3">DKOM unlinking</td></tr>
<tr><td>ObjectDirectory, DRIVER_OBJECT</td><td>ZwQueryDirectoryObject</td></tr>
<tr><td>Service Record List,<br>SERVICE_RECORD</td><td>EnumServicesStatus</td></tr>
<tr><td>List of kernel mode system threads,<br>ETHREAD</td><td>Volatility's orphan plugin</td><td rowspan="2">DKOM patching</td></tr>
<tr><td>MmUnloadedDrivers,<br>UNLOADED_DRIVERS</td><td>• ReadDebuggerData<br>• Volatility's unloadedmodules plugin</td></tr>
<tr><td>Service Record List in registry</td><td>RegEnumKeyEx</td><td>Removing registry keys</td></tr>
</table>

Two signature types can be distinguished:

- Signatures based on short byte sequence also known as pool-tag scanning;
- Signatures based on stand-alone bytes also known as magic numbers in the structures.

**Pool-tag scanning**. This signature is based on the fact, that each structure from one list contains the same four byte tags. This is one of the most popular ways to find structures in the memory. This tag could be added to the structure using two ways: with the function ExAllocatePoolWithTag or manual by a programmer. In the first case a four-byte tag value is added automatically at the beginning of the allocated pool memory. These values are reserved for each system structures. The pool tag list of Windows drivers is given in Rhee, (2009). In the second case such byte signature is added by the programmer. For example, while a driver is started, SCM creates the SERVICE_RECORD structure and adds the pool-tag value: (*ServiceRecord)->Signature = SERVICE_SIGNATURE, where SERVICE_SIGNATURE is 0x76724573 or "sErv" in ASCII. The details of this manipulation are in the source code (Microsoft, n.d. d). M. Ligh used the pool-tag scanning detect Windows threads (Ligh, 2011).

**Magic numbers in the structures**. The second type of signature is based on the fact that structures from one list have some common peculiarities. This signature includes only the bytes, whose values are the same for all linked structures on the list. Also apart from the single bytes, we can also use the fact that kernel mode structures include the fields, whose values are linked with other kernel mode structures. Therefore these values exceed the values of 0x8000_0000 (Schuster, 2006). This peculiarity was described by (Tsaur, & Chen, 2010; Haukli, 2014) and after that was enhanced in Dynamic Bit Signature (DBS) to detect hidden structures (Korkin, & Nesterov, 2014).

The summary table of signature-based search methods of driver structures is in Table 2.

As a result, simultaneously applying both anti forensic techniques: DKOM unlinking and DKOM patching renders drivers' structures useless for detection and significantly hinder further analysis. Next, we will cover driver detection approaches, which do not rely on driver structures; therefore, these approaches are able to detect the hidden driver, which uses both the above mentioned countermeasures.

Table 2 Summary table of signature-based search methods of driver structures

| List and structure names | Signature examples | Signature using example |
|---|---|---|
| PsLoadedModuleList, KLDR_DATA_TABLE_ENTRY | Tag signature "MmLd" | Cohen (2014) |
| Service Record List, SERVICE_RECORD | Tag signature "sErv" or "serH" | Hidden Service Detector (hsd) by EiNSTeiN_ Ligh, Case, Levy, & Walters (2014f)<br><br>Volatility's *SvcScan* plugin by Ligh, Case, Levy, & Walters (2014g) and Vomel, & Lenz (2013) |
| ObjectTable, DRIVER_OBJECT | Tag signature "Driv" | Volatility's *modscan* plugin by Ligh, Case, Levy, & Walters (2014h) |
| | Byte signature | Tsaur, & Chen (2010) and Haukli (2014) |
| | Dynamic bit signature | Korkin, & Nesterov (2014) |

### 2.4. Signature-based Detection of Drivers Files

Signature-based search of drivers' files uses priority known fragments of drivers' files as signatures; therefore, these methods are not susceptible to manipulation with drivers' structures. In comparison with the previous signature-based methods, these methods use a similar byte-to-byte search for drivers' files, but not for drivers' structures.

There are three types of signatures to find driver files to be discussed:

- ASCII strings;
- Magic numbers in the PE-header;
- OpCode patterns.

**ASCII strings** The first type of signatures includes various strings from an executable file. The best known example of ASCII strings signatures is "This program cannot be run in DOS mode" from MS-DOS header (Timzen, 2015; x86 Disassembly/Windows Executable Files, n.d.). M. Russinovich used this string while analyzing Stuxnet (Russinovich, 2011). Also, import table, in other words, symbolic names of functions belonging to ASCII strings signatures. The example of this signature is used to analyze packed PE files, "The Presence of certain functions in the import" (NTInfo, 2014).

**How to bypass ASCII strings signatures?** The ASCII strings signatures are not resilient to the following countermeasures. First the hidden driver can overwrite its MS-DOS header in the memory; and this manipulation does not influence its OS work (Qark, n.d.). Moreover, to hinder detection a hidden driver can use only two imported functions LoadLibrary and GetProcAddress. It is also able to perform the function-resolution tasks and to call on any other functions (Eagle, 2011). As a result, the import table will include just only two names, which will not be enough for detection.

**Magic numbers in the PE-header**. The second type of signature is based on the peculiarities of the PE-header content (ELF-header in UNIX case) or in the magic numbers. This signature is a byte template, which includes the field values of PE header and their corresponding offsets from the beginning of the file (Dorfman, 2014; Choi, Kim, Oh, & Ryou, 2009). An example of the application of magic numbers in the PE-header to detect unknown malware is presented by Wang, Wu, & Hsieh (2009). An example of adopting this signature to ELF-header for unknown malware detection in the OS Linux is given in Farrukh, & Muddassar (2012).

**How to bypass magic numbers in the PE-header?** The hidden driver can bypass this type of signature using a similar overwriting technique. After the driver has been loaded its PE-header integrity is not needed. Therefore a rootkit can overwrite the content of PE-header and bypass signatures based on magic numbers. This approach was also proposed by Tsaur, & Wu (2014c).

**Opcode patterns**. The third type of signatures is based on the common factors in the content of the compiled file. Each driver is an executable file, which includes sets of machine language instruction or operation codes, and further opcodes, which are executed by CPU. An Instruction Set Architecture (ISA) includes a specification of the set of opcodes. Different drivers have similar opcode fragments, because they are executed on the same CPU architecture. The most popular example of using opcodes as a signature is the search of starting and ending of functions. The corresponding opcodes are also known as functions prolog and epilog and they depend on calling convention (Calvet, 2012). There are three popular calling conventions that are used with C/C++ language: stdcall, cdecl and thiscall (Chen, 2004). Opcode based signatures are used to detect various malware (Santos, Brezo, Nieves, Penya, Sanz, Laorden, & Bringas, 2010; Santos, Sanz, Laorden, Brezo, & Bringas, 2011; Zolotukhin, & Hamalainen, 2014). There are numerous examples of using opcodes for detection (Ghezelbigloo, & VafaeiJahan, 2014; Singla, Gandotra, Bansal, & Sofat, 2015; Yu, Zhou, Liu, Yang, & Luo, 2011).

An example of using opcode byte combination to find hidden service structures is given in (Ligh, M. 2015a; Ligh, M. 2015b). Opcode patterns are also used in firmware analysis. The pattern matching tool Binwalk detects the potentially executable data by identifying known function prolog and epilog patterns (Binwalk, 2015).

Opcode-based signatures could be further developed using control flow analysis (Ding, Dai, Yan, & Zhang, 2014) and data mining techniques (Siddiqui, Wang, & Lee 2008; Santos, Brezo,

Ugarte-Pedrero, & Bringas, 2013) and also varied classification techniques (Shabtai, Moskovitch, Feher, Dolev, & Elovici, 2012).

**How to bypass opcode patterns?** Several authors have pointed out that applying prolog and epilog signatures as well as using other ISA opcode patterns does not make it resilient to anti-forensic techniques. A rootkit can bypass opcode patterns by obfuscating prolog and epilog functions. An example of hiding functions call is given in Emil's Projects, (2010). Also malware can evade the signature based detection by packing the original code using custom packers (Arora, Singh, Pareek, & Edara, 2013).

As a result, we can conclude, that despite the fact that the signature-based search of drivers files is resilient to manipulation with driver structures, this search is susceptible to other anti-forensic techniques. An intruder can circumvent signature based search of drivers files in memory by overwriting sensitive data for ASCII-based signatures and signatures based on PE-header features; as well as encrypting, packaging and obfuscating of content of a driver file (Saleh, Ratazzi, & Xu, 2014; Dang, Gazet, Bachaalany, & Josse, 2014; Aronov, 2015).

Further, we will present an analysis of driver detection approach, which is resilient to combination of all previous anti-forensic techniques.

### 2.5. Statistical Detection of Driver Files

Statistical detection of driver files is able to reveal loaded drivers in the memory files, without using any fixed signature. This detection approach is based on the fact that the content of driver files is significantly different from the content of other memory parts.

Statistical detection of driver files includes two phases. In the first step, we evaluate the memory content by calculating various statistics for each memory offset. This technique is also known as the sliding windows approach. In the second step, we find memory regions with abnormal values for the calculated statistics. These memory regions are very likely to have a driver code or raw machine code.

An impulse of using the statistical memory analysis to find an executable code in the memory is given in Lyda & Hamrock, (2007), which specified the Shannon's entropy notion and was the first to apply binary entropy. The authors discovered that by comparing entropy values it is possible to separate different data types: plain text, native executables, packed executables and encrypted executables (Brown, 2009).

According to the blog post (Suszter, 2014) "entropy analysis is very useful to locate compressed, encrypted, and other way encoded data; higher entropy can indicate encoding of some kind; lower entropy is likely to include anything else such as text, code, header and data structures."

Our analysis of publicly available sources shows that there are no facts of using entropy to detect hidden drivers in the memory. However, entropy is applied in various spheres of cyber security (Matveeva, 2014), such as finding an executable code in the office document files (Jochheim, 2012; Pareek, Eswari, Babu, & Bangalore, 2013; Iwamoto, & Wasaki, 2014; Tabish, Shafiq, & Farooq, 2009), in PDF-files (Schmidt, Wahlisch, & Groning, 2011; Pareek, Eswari, & Babu, 2013), in the network traffic (Nguyen, Tran, Ma, & Sharma, 2014), in the analysis of unknown binary files (Yurichev, 2015). Entropy analysis is also used to find cryptographic keys in the memory dump (Maartmann-Moe, 2008). This successful experience can be applied to find hidden driver content in the memory and detect rootkits.

Entropy analysis of individual files is the most commonly used method to solve the task of files classification, for example, to separate packed and encrypted file from an unprotected file. According to (Devi, & Nandi, 2012) 'it is very important to figure out whether a given executable is packed or non-packed before determining whether it is malicious or benign'. If certain conditions are true, such as, the file has a high entropy value, unknown MD5 hash value and also it has no digital signature, it is most likely that this executable file is malware (SANS Institute, 2014; Nataraja, Jacob, & Manjunatha, 2010). In addition, in this author's opinion antiviruses can usually detect weird applications using entropy. To develop a 'well done rootkit' we need to be creative about entropy (Lupi, 2011).

The sliding window approach allows us to calculate a variety of statistics. Together with the formula of binary entropy by (Lyda, & Hamrock, 2007) there are other ways of calculating entropy (Hall, & Davis, 2006). Also apart from entropy in the sliding window approach, we can use other statistics such as arithmetic mean, Chi square and Hamming weight (Conti, et al., 2010). Jochheim, (2012) thinks that the most useful statistics in file classification are the following: average, kurtosis, distribution of averages, standard deviation and distribution of standard deviations. One of the prospective methods to measure binaries is to apply Kolmogorov complexity for detecting malware (Alshahwan, Barr, Clark, & Danezis, 2015).

In the case of memory dump, we do not have separate files and so we can apply the sliding window approach or its various modifications. One such example is given by (Matveeva, & Epishkina, 2015). The authors proposed constructing a byte value / byte offset dependency graph. Byte offsets are placed on the horizontal axis, and byte values are on the vertical axis. After that, the authors calculate and construct a frequency byte distribution for various window-form fragments of the first plot. The authors suggest using the following fragment sizes 100x100 and 2000x50.

Another approach is to use the following configuration of sliding window method: each block has fixed length 256 bytes, "Each block has an overlap of 4 bytes." "A short-term Fourier analysis is applied to every 4 bytes of the entropy stream." (Jochheim, Schmidt, & Wahlisch, 2011)

For further analysis of the memory dump these authors suggest the following methods: Short-term Fourier Transform (STFT) (Schmidt, Wahlisch, & Groning, 2011; Iwamoto, & Wasaki, 2014) or Wavelet transform (Matveeva, & Epishkina, 2015).

Other ways of calculating statistics apart from binary entropy by are suggested by (Lyda, & Hamrock, 2007), who using one byte, we can also calculate entropy using two bytes or using bigram analysis (Nataraja, Jacobb, & Manjunatha, 2010), as well as using three bytes or trigrams (Conti, et al., 2010). Along with the analysis of individual values statistics, we can also evaluate the memory content using n-gram analysis (Pareek, Eswari, & Babu, 2013; Nath, & Mehtre, 2014).

It is possible to carry out memory analysis using both manual and automatic modes. The idea of automated classification of different file types, such as random, text, machine code, bitmap, compressed images, encoded and encrypted data exposed in the paper (Conti, et al., 2010). To solve a classification problem the authors first calculated threshold values for different data types by using test samples. Their results solved the classification problem with the appropriate values of false positives and false negatives. An alternative example of solving the classification problem by using entropy is given in (Bat Erdene, Kim, Li, & Lee, 2013).

Artificial intelligence methods look very promising for solving the classification problem, but the authors thought that these methods are insufficiently precise, because of errors of the first and second types (Jochheim, 2012).

To make manual analysis of memory easier and faster we can use various visualization techniques. In his paper Jocheim suggested using the following byte blot: each byte in the file was colored according to this bytes hex value, e.g. zero bytes and bytes with value 0xFF are black and white correspondingly. Using this approach, we can easily locate zero pages in the memory as well as the particular sections of an executable file – .text, .data, .rsrc.

An alternative approach for visualizing data uses space-filling curves proposed by A. Cortesi. He analyzed various approaches of how to visualize the file content, such as Zigzag, Z-order and Hilbert to find encrypted blocks (Cortesi, 2012). As a result of this work the author developed tools to visualize content using both web-site (Cortesi, 2015a) and local host PC (Cortesi, 2015b).

Visualization of files or memory dumps is an up and-coming direction in cyber-security, which was, confirmed by various projects: thus the cantor.dust project was presented at the REcon'13 conference, Binwalk, binglide, Vix/Biteye, senseye and many others (Visualizing ELF binaries, 2014).

In addition, it is possible to apply a combination of the approaches mentioned before. An example of using signature-based and statistical approach was presented by (Merkel, Hoppe, Kraetzer, & Dittmann, 2010).

**How to bypass statistical detection?** At the same time, the statistical detection approach is also vulnerable to anti-forensic techniques. The idea of countermeasures is to apply the manipulations, which decrease the entropy of driver content and as a result seriously hinder localizing a driver code in the memory areas.

There are only two publicly available countermeasures to decrease entropy: the first is to use Multiple Files and the second is to add memory blocks with zero entropy.

The first method is used in the Stuxnet and Flame, which deliberately decreased the entropy values using Multiple Files to achieve ‘lesser maliciousness entropy’ (Teller, 2013; Schuberth, 2014).

The second one is also known as a meta obfuscation technique by null content insertion. Executable code can also hide itself by reducing entropy values of its memory content with the help of including memory fragments with low entropy. An example of such manipulation is used in spyware program Zeus and it includes inserting blocks of symbols with zero entropy. Moreover, according to the blog post, many packers use this anti-forensic technique to hide the fact that this file has been encrypted – “some very good packers and protectors of malware try to reduce entropy by inserting zero bytes in data. The reason is that virus scanners react to files with high entropy” (NTInfo, 2014).

To bypass the circumvention, i.e., to detect such malware the authors (Pedrero, Santos, Sanz, Laorden, & Bringas, 2012) proposed a way to localize and delete such data blocks and after that calculate entropy values. The latter was made possible by applying byte histograms to analyze the contents of the file. However, applying byte histograms is also vulnerable to the corresponding anti forensic technique, as discussed below.

### 2.6. Conclusion

The above analysis shows that all popular approaches to detect drivers are susceptible to anti-forensic techniques, see Table 3.

By joining the results of this chapter and the work (Korkin, & Nesterov, 2014) we can state that the most popular framework for memory analysis, Volatility is also vulnerable to anti-forensic technique and moreover that malware can overcome Volatility in both stages: memory acquisition and memory dump analysis.

To sum up we can conclude that a driver can be hidden in different ways (Jason, 2012) but the results will always be the same – running uncontrolled code in the privilege memory area and the content of this code will have a low entropy value.

This work focuses on the anticipatory development of advanced cyber-security solutions. As a result, in this project we will create a new hidden driver detection method, which will be resilient to current driver countermeasures and their possible development.

The task to detect a driver comes down to the recognition task between the executable driver code and data fragments, which are stored in the memory.

Table 3 The list of popular approaches to detect drivers and their vulnerabilities

| **The ways to detect drivers** | | **Anti-forensic technique** |
|---|---|---|
| Using driver structures | Using links between structures | DKOM unlinking |
| | Signature-based search | DKOM patching |
| Using content of driver files | Signature-based detection | DKOM patching & PE packing |
| | Statistical-based detection | Memory patching |

## 3. THE IDEA OF THE HIGHEST STEALTH MALWARE (HIGHSTEM)

In this section we will look at a prototype of the most hidden driver or Highest Stealth Malware (HighSteM). The goal of this section is to create a list of possible anti-forensic techniques, which include both evasion mechanisms for popular detection approaches, and the present authors' ideas on how to avoid statistical-based detection (Pedrero, Santos, Sanz, Laorden, & Bringas, 2012). We are trying to create anti-forensic measures, which can prevent detection even by future detection tools. As a result, we will formulate the most difficult scenario for detection. The next section will deal with detection of HighSteM.

Malware can use different ways to start up: using shellcode in PDF-file or by registry facilities for loading instead of the file system (Santos, 2014; Marcos, 2014), but all in all the malware executable code will be loaded and reside in the memory.

It is well known that the code, which is running in the lowest level, and is close to hardware, has the greatest potential for self-concealment There are various levels of code execution: inside OS user mode (ring 3), kernel mode (ring 0) and outside OS VMX root mode or hypervisor (ring -1), SMM (ring -2) and AMT (ring -3). We chose kernel mode (ring 0), which is the most popular among malware and spyware platforms. HighSteM will be the kernel mode driver for Windows OS or a loadable kernel module (LKM) for UNIX-based systems.

As a basis to develop the HighSteM prototype, we will select the FU rootkit (Blunden, 2012) or a stub driver, which is loaded using ATSIV utility by Linchpin Labs (Linchpin Labs, 2010). The results of our preliminary research revealed that ATSIV uses an undocumented startup method, which conceals a driver from the most popular anti-rootkit tools (Korkin, 2012). In addition, it is possible to use Turla Driver Loader, which loads a driver without involving Windows loader (hfiref0x, 2016)

We can load any driver using both built-in Windows tool (eg Service Control Manager) and by third party software (like Atsiv or Turla Driver Loader). In the first case drivers information is added in all system lists, while in the second case just a few lists will be updated.

The HighSteM prototype will apply a variety of anti forensic techniques. First of all HighSteM will include techniques to prevent the detection of existing approaches, the details are in Section 2 Then we present some ideas on how to improve these anti-forensic techniques. HighSteM could improve Zeus's manipulations to hide from entropy analysis by the following three steps:

1. Insert blocks of symbols, with low nonzero entropy value;
2. Use blocks with different size;
3. Significantly increase size and number of inserted blocks.

On the one hand applying these three steps makes a driver definitively hidden from Pedrero's methods (Pedrero, Santos, Sanz, Laorden, & Bringas, 2012) and, moreover, they can help to develop the most complex case for detection,. However, on the other hand, these steps will lead to certain negative effects: the size of the driver will increase and its speed of operation will decrease. The reasons for these disadvantages are blocks of inserted symbols, so there will be many time consuming jumps between pieces of the driver code.

We also propose an idea based on HighSteM's payload and how to realize this in a stealthy way. The main strategy is to elicit sensitive information by reading various memory regions, without using the functions, which can be hooked by anti-viruses. An idea of the keylogger, which collects keystrokes by checking the corresponding memory fragment, was first proposed by Ladakis et al (2013). These authors apply the GPU facilities to access physical memory. They also proposed to check a memory region by including keystrokes with 100ms delay. This time is enough to collect all keystrokes with an average typing speed and also with optimal system's overhead, which is less than 0.1%. This paper focused only on the Linux OS.

Adapting this idea for Windows OS is given by (Stewin, & Bystrov, 2012). These authors describe how to find the OS structure that contains the most recent keystrokes, but without its internal details. To find out these details the authors propose using reverse-engineering analysis of the kbdhid.sys file, which may be challenging.

Our preliminary research reveals, that keystrokes buffer is stored in the DEVICE_EXTENSION structure, the source code for a USB keyboard is here (Microsoft, n.d.-e) and also for a PC/2 keyboard (Microsoft, n.d.-f). This structure includes the KEYBOARD_ATTRIBUTES structure, which contains the desired codes of keystrokes and other additional flags. The similar structure – KBDHID_DEVICE_EXTENSION is used in ReactOS, which is close to Windows OS (ReactOS, n.d.).

This memory-based monitoring technique looks promising. On the one hand, it acquires sensitive information and on the other hand, cyber-security tools and well-known event tracing solutions are not able to control memory access. This is the most difficult situation for detection.

In the next section, we suggest some ideas of how to detect HighSteM-based rootkits.

## 4. DETECTION OF HIDDEN SOFTWARE UNDER COUNTERMEASURES

To achieve this goal the following three tasks need to be tackled:

A. Research and develop a prototype of the hidden driver, which can overcome existing detection methods.
B. Check the driver prototype using existing detection methods and tools.
C. Design new methods to detect hidden drivers using analysis of Windows OS memory.

To solve task A we are going to use the example of the keyboard driver filter as the basis for the prototype of the hidden driver (Blunden, 2009). We are planning to take the following measures to hide this driver. We will load the driver prototype with the help of publicly available ATSIV utility by (Linchpin Labs, 2010). This driver will be concealed from the signature search by patching its PE-header in the memory. To develop HighSteM we will solve the optimization problem, with the following two variables – the size and number of inserted blocks of symbols and the following three constraints:

- The maximum driver size – up to 10 megabytes;
- The level of decreasing speed of operation – no more than two times;
- The level of entropy decreasing – no less than two times.

To speed up the development and testing of this driver we will use the already prepared memory analysis system (Korkin, & Nesterov, 2014).

To solve task B we will use the following popular tools to detect a deliberately hidden driver: Kaspersky TDSSKiller, GMER, RootRepeal, Avast Anti-Rootkit, Dr.Web CureIt!, Sophos Anti-Rootkit, F-Secure Blacklight and the most popular memory analysis platform – Volatility Framework (MidnightCowboy, 2015; Ligh, Case, Levy, & Walters, 2014). As a result, we will experimentally prove that these cyber security solutions are not able to detect this prototype of the hidden driver.

On stage C, we are going to apply these three ideas to find the executable code in the memory dump:

1. New methods of entropy calculation and data analysis:
   1.1. Use function P*(2-P^2) instead of P*log(P) to calculate entropy. Our preliminary analysis showed that this function grows faster on the interval from 0 to 1, than the original entropy and that is why it looks more appropriate for the analysis of computer memory;
   1.2. Analyze dependence of calculated entropy values from different lengths of sliding windows and its further spectral and wavelet analysis;
   1.3. Apply methods of digital photogrammetry to find the executable code in the graphical diagrams of the memory content;
   1.4. Use Zipf–Mandelbrot law to analyze the executable code and data in the memory content.
2. Disassemble memory content and carry out the thorough analysis of the received assembler code, using the following ways:

2.1. Instruction Frequency Analysis;
2.2. Evaluation of the logic of the assembler code;
2.3. Design and analysis of the control flow graph based on the received set of the assembler code.

To implement the first idea, we are planning to use numerical computing environment MATLAB. Applying MATLAB built-in functions helps us to focus on statistical ideas rather than on their implementation and testing.

Methods of digital photogrammetry are commonly used to locate different objects on the digital images, for example, human faces on photographs. When it comes to hidden drivers detection we have to face a similar challenge – to localize memory fragments with anomalous entropy values, which correspond to the executable code (Cortesi, 2012; Kohli, Lempitsky, & Barinova, 2015).

The Zipf–Mandelbrot law is used to check the self-organization, correctness and systematicity of literary works. The executable code has similar properties, which is why we will use this law to evaluate the code and data in the memory.

To realize the second idea about applying disassembling, we will use the Capstone and BeaEngine, which are the open source libraries to disassembly both 32 and 64 bits code (Nguyen, 2014).

Analysis of frequency of assembler instructions includes preliminary and detection phases. In the preliminary phase, we will calculate the threshold frequency of instructions from memory fragments, which include only the executable code and only the data. In the detection phase, we will calculate frequency instructions from each memory fragment and compare their frequency with threshold values. Evaluation of logic will be made by checking that the result of the instructions execution does not overwrite the previously achieved results. The design and analysis of control flow graph will be performed using Interactive Disassembler IDA (Blunden, 2012). A similar idea of disassembling the part of the code is utilized in Volatility's *driverirp* plugin with '--verbose' flag to reveal TDL3 infection. TDL3 uses the Stealth Hooks technique and Redirector Stubs to hide the fact of hooking functions (Ligh, Case, Levy, & Walters 2014i).

We will consider the following two limitations of this research:

1. Analysis of the memory content will be given only for 32-bit and 64-bit Windows 7 OS, as the most popular OS, without analysis of Unix-based OS, such as Mac OS, Linux and mobile OSes.
2. We will consider the executable samples, which do not apply obfuscation and polymorphism techniques.

However, such analysis requires significant computing capabilities from the analyst's workspace, for example, rootkit detection system MASHKA spends about half hour to check memory dump, which is far too long and hence not applicable in practice.

To speed-up the analysis we proposed an idea based on the use of modern video or graphics cards. To do this we will present an idea of accelerating memory analysis using the CUDA library and NVIDIA graphics card.

## 5. GPU ACCELERATED SIGNATURE-BASED MEMORY DUMP ANALYSIS

Nowadays modern workstations, laptops, notebooks and even tablets are equipped with one or several GPU components. Besides gaming opportunities such components offer general high performance parallel computing possibilities for a long time. For memory dump analysis it is important that such devices can transfer CPU load to GPU, and also are very suitable for signature-based and statistical analysis. Obtaining memory dump on modern workstations is a separate CPU-intensive task. Thus permanent analysis is barely possible but periodic continual dumping is a realistic scenario. Such time-based periodic dumping and analysis can dramatically load CPU and may be inappropriate during simultaneous regular work.

This paper offers a statistical confidence estimation of hidden malware and several algorithms to form a criterion of significance for memory dump segments. Most of the mentioned algorithms may be efficiently run in parallel due to inner data parallelism. As a result, most of the analysis may be effectively transferred to GPU dramatically diminishing CPU-load during processing. This section also describes the principles of fine-grained algorithm splitting to run memory dump analysis effectively on hybrid CPU/GPU aware architectures.

The reason we need to develop an appropriate GPU kernel based code for CUDA aware GPUs is the internal highly localized cohesion between memory dump data segments and applicable statistical algorithms. Such cohesion makes memory dump analysis very close to classical box filtering and convolution filters run on GPU. Both of them have good and effective implementations for CUDA aware GPUs and also show very high scalability and performance. Memory throughput is no longer possible because of a bottleneck due to extensive data transfer between host based and GPU-based memories. As a part of the newly announced “Boltzmann Initiative” AMD presents the Heterogeneous-computer Interface for Portability (HIP) tool. New heterogeneous system architecture will allow us to automatically convert CUDA code by HIP and expand the possible hardware base available to run what was formerly an exclusively CUDA-based applications (Silcott, & Swinimer, 2015). So nowadays CUDA may be the best choice to meet current and future requirements for consumer and enterprise-based hardware from both vendors.

This approach further involves differentiating memory dump analysis algorithms based on different memory-access profiles during processing. Two main profiles are easily discovered: local signature-based detection and memory lookups to resolve virtual to physical memory layouts (Korkin, & Nesterov, 2014).

Local signature based detection is the most appropriate for running on GPU; memory model and kernel execution principles are best suited for running such analysis. CUDA box filtering and convolution based examples bundled with the CUDA toolkit are optimal starting points, and as such processing is effectively done to boost filtering and analysis performance.

To utilize multiple available GPU there are two ways to handle them either as independent devices or alternatively as a single one through a unified memory model using modern CUDA improvements. Starting with CUDA 4 unified virtual addressing is supported, and this provides a single virtual memory address space for all the memory available in the system. Memory addressing and management enables pointers to be accessed from the GPU code no matter where in the system they reside, whether in device memory (on the same or a different GPU), host memory, or on-chip shared memory. Latter improvements to overcome PCI-Express’s low bandwidth and high latency are available since CUDA 6 platform. This brings out managed memory, which is accessible to both the CPU and GPU by using a single pointer. The key of the improvement is that the system automatically migrates data allocated in unified memory between the host and device so that it looks like CPU memory to code running on the CPU, and like GPU memory to code running on the GPU.

Despite the above mentioned advances it is a complicated task to utilize such functionality when non-uniform GPUs are installed and so hand-coded algorithm splitting must be done to avoid non-local memory addressing. To counter this another approach was developed with high-grained algorithm splitting based on memory-access patterns during signature-based detection.

### 5.1. Hybrid CPU/GPU Architecture Requirements

Software architecture for signature-based detection must allow effective highly scalable and accelerated memory dumps analysis. Memory datasets may be accessed through interposes - communication with dumping routines or inner communication through shared thread storage. Another source of bulk datasets may be from some SOCKET-based network layer where datasets from different nodes are aggregated and analyzed. As a part of the research such a SOCKED-based layer was developed to further allow research and investigation into building trusted networks with secure runtime memory dump analysis and incident systems.

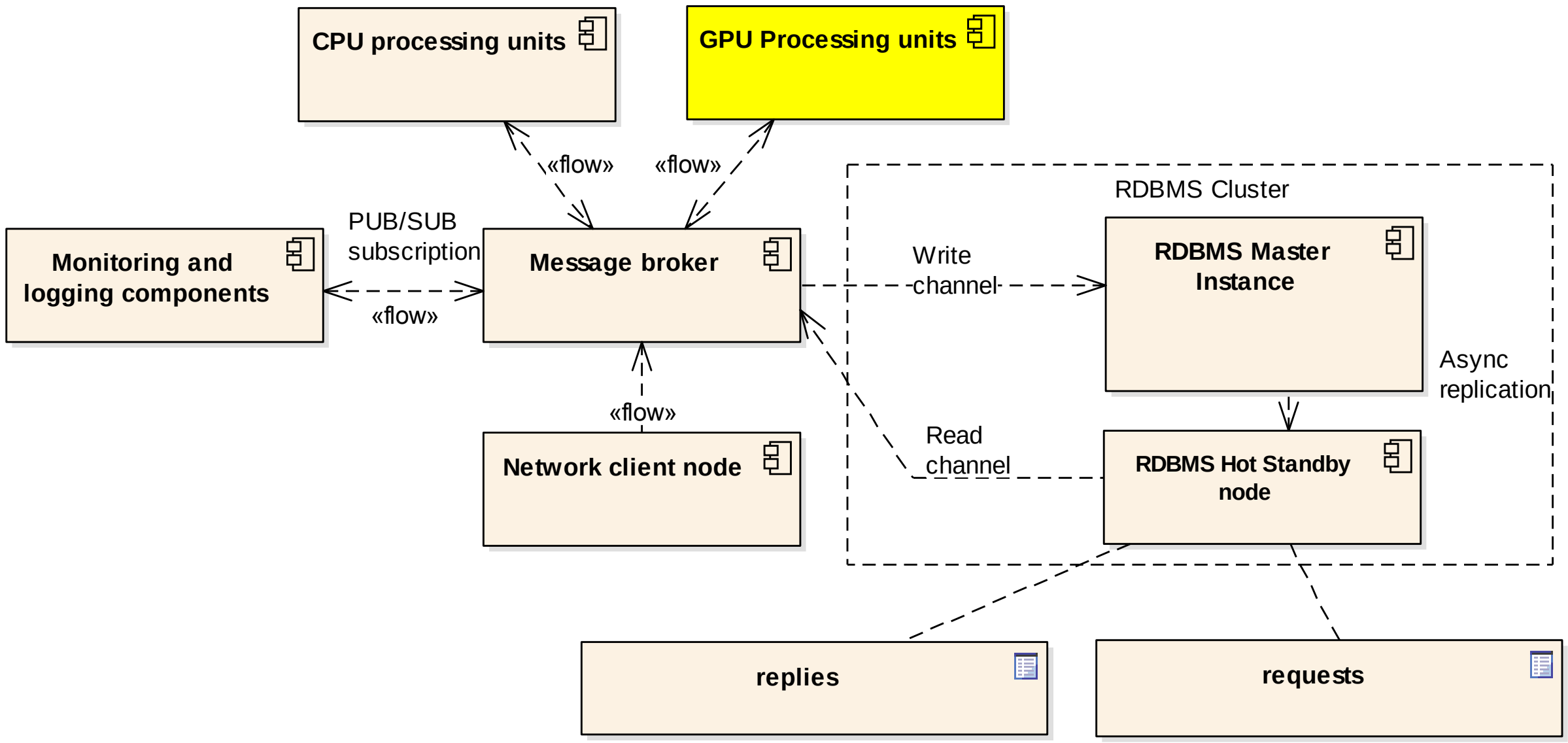

Figure 3 Architecture of a network-aware software platform for memory dump analysis

A signature-based threaded detection system must:

- Effectively run on nodes with single or multiple GPUs, on multicore and multiprocessor systems;
- Allow different GPU memory layouts and data handling strategies to allow benchmarking and work on different GPU architectures;
- Meet CPU counterpart implementations;
- Do benchmarking and algorithm variations for different global, texture, constant and shared memory available resources and data arrangement;
- Minimize thread divergence, where threads of the same wrap branch to different section of code.
- Confirm single instruction multiple thread manner and prevent GPU scheduler from allocating extra execution slices to possible branches;
- Ensure coalesced memory aces by wrapping branches.

The goal to promote a network-aware software platform to memory dump analysis naturally leads on to the development of scalable tool and architecture to be able to simultaneously perform statistical processing and provide analytics. The main architecture principle lies in modular and an extensible object-oriented pipelined framework to allow parallel code execution on shared memory multi-processor platforms. A reliable network aware software platform consists of a dedicated message broker and dedicated RDBMS cluster to register network connections and analysis requests, as shown in Figure 3.

## 5.2. Overview of Network-aware Architecture

A server-side component was also developed. It provides endpoints of communication using a socket network programming interface. Server-side components are responsible for handling connections from the different nodes, which make requests to perform the forensics analysis of their memory dumps.

In its turn an object-oriented pipelined framework is responsible for promoting event demultiplexing and concurrent processing of the heterogeneous CPU/GPU architecture. The application starts a stream that presents a set of hierarchically-related analysis and reporting services. Network data demultiplexing and further asynchronous processing Reactor (Reactor pattern, n.d.) was selected as the main programming pattern. This programming pattern makes a unified asynchronous event handling which is generated by a timer-driven callout queue, I/O events received on communication ports, events from GPU kernels and CPU-processing threads, as shown in Figure 4.

Each processing step in the hierarchically-related analysis and reporting stream presents a threading pool that makes parallel processing possible. The thread pool is configurable and responsible for spawning, executing, synchronizing and gracefully terminating managed threads and data flow through the processing stream. Initial benchmarking is done to adjust the different task decomposition policies to permit parallel processing on GPU and CPU resources with different computing capabilities and performances.

### 5.3. Overview of Pipelined Processing Workflow. Further Development

Network-aware software platforms for memory dump analysis may be an essential component for Intrusion Detection Systems (IDS). Such components may reside on dedicated high performance hardware or virtualized platforms. High performance tools that can utilize available multicore CPU and multiple GPU are required. Such tools process highly sensitive data and further research and development must be done to assure security and confidence. Digital signing and correct identification, authentication and authorization schemes must be provided and there is also a need for mechanisms for integration with the latest directory services.

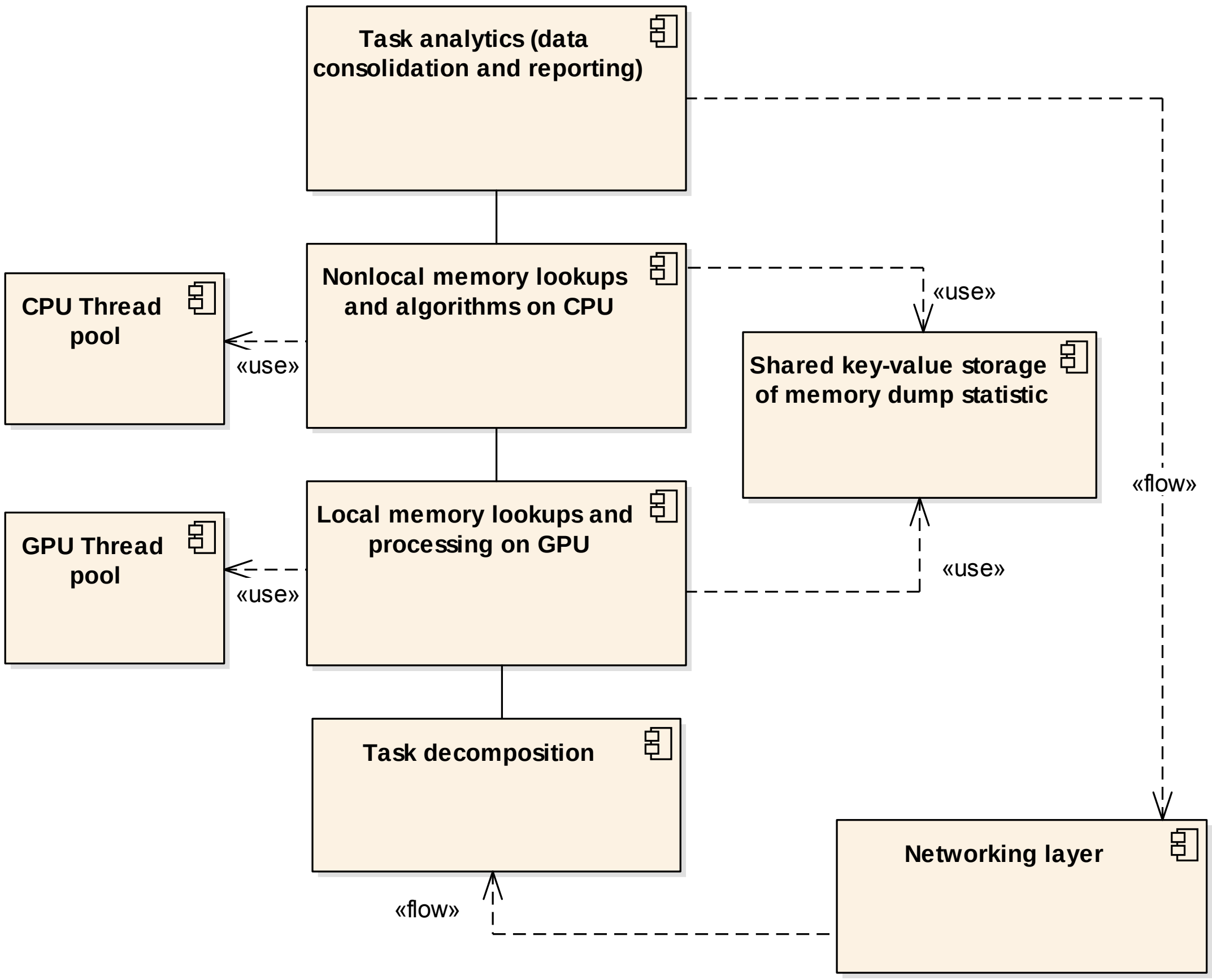

Figure 4 Architecture of Parallel Processing on GPU and CPU resources

## 6. CONCLUSIONS & FUTURE WORK

1. Detection of zero-day malware is a well-known and current central challenge in cyber security. This analysis shows that existing detection approaches are susceptible to rootkits which apply a variety of different countermeasures.
2. To create the most difficult case scenario for detection we propose a driver prototype based on existing anti forensic techniques and their possible upgrades. To find such a driver we propose various improvements on statistical-based detection.
3. We proposed an idea of using the facilities of both GPU and CPU to speed up memory analysis.

### 6.1. The Idea of a Dynamic Memory Map

Another idea of rootkit detection is the Dynamic Memory Map tool. This tool will help to visualize memory content of virtual or physical addresses with multicolored rectangles, corresponding to different drivers and kernel services, as well as a user mode process. If we have found a memory fragment with an executable code, which is not registered in the OS, it will mean that our OS has been infected and this executable code has to be subject to an additional inspection. This tool will have two modes of operating: as an on-air map and a logger of all memory accesses to/from the chosen memory scope. Using Intel VT-x with Extended Page Tables (EPT) technology we will be able to achieve portability for all new Intel CPUs, reduce significant performance losses and make this tool resilient to common OS-based anti-forensic techniques. A collaboration with Satoshi Tanda (Tanda, 2015) has suggested a preliminary implementation of such a tool and early results look promising. This results will be presented at the REcon conference 2016 in the paper “Monitoring & controlling kernel-mode events by HyperPlatform”.

### 6.2. Applying Virtual Reality (VR) Headset to Digital Forensics Applications

Modern technologies present new devices for head-mounted display (HMD) system: Virtual Reality (VR) headset or VR gear, which are extremely popular nowadays. These devices have the following advantages:

- Comfortable for eyes, “it's like watching a 130-inch television screen from 10 feet away”;
- Cheap, “Google Cardboard headsets are built out of simple, low-cost components”;
- VR useful in a wider variety of roles, not only for gaming.

We propose an idea to use the opportunities of VR headset in an incident response during the analysis of various system logs and memory dumps. In addition, VR headset can be used to analyze the control-flow-graph of disassembled code during reverse-engineering process. New 3D view from VR headset, instead of existing 2D picture from desktop PC, could speed-up analysis and makes it more easily.

## 7. ACKNOWLEDGEMENTS

We would like to thank Andrey Chechulin, Ph.D research fellow of Laboratory of Computer Security Problems (Scientific advisor – Prof. Igor Kotenko) of the St. Petersburg Institute for Informatics and Automation of the Russian Academy of Science (SPIIRAS) for his insightful comments and feedback which helped us to improve the quality of the paper substantially.

We would like to thank Samantha Gonzalez from Miami High School, Florida, US and Sarah Wilson teacher of English, Kenosha, Wisconsin, US for their time and effort in checking a preliminary version of this paper.

We would also like to thank Ben Stein, teacher of English, Kings Education, London, UK for his invaluable corrections of the paper.

## 8. REFERENCES

[1] Alshahwan, N., Barr, E.E., Clark, D., & Danezis, D. (2015). Detecting Malware with Information Complexity. [arXiv preprint]. Retrieved on December 8, 2015, from http://arxiv.org/pdf/1502.07661.pdf

[2] Aronov, I. (2015). An Example of Common String and Payload Obfuscation Techniques in Malware. Security Intelligence. Retrieved on December 8, 2015, from https://securityintelligence.com/an-example-of-common-string-and-payload-obfuscation-techniques-in-malware/

[3] Arora, R., Singh, A., Pareek, H., & Edara, U.R. (2013). A Heuristics-based Static Analysis Approach for Detecting Packed PE Binaries. International Journal of Security and Its Applications. 7(5). 257-268. Retrieved on December 8, 2015, from http://www.sersc.org/journals/IJSIA/vol7_no5_2013/24.pdf http://dx.doi.org/10.14257/ijsia.2013.7.5.24

[4] Bat-Erdene, M., Kim, T., Li, H., & Lee, H. (2013). Dynamic Classification of Packing Algorithms for Inspecting Executables using Entropy Analysis. Proceedings of the 8th International Conference Malicious and Unwanted Software: "The Americas" (MALWARE). Fajardo, PR. http://dx.doi.org/10.1109/MALWARE.2013.6703681

[5] Binwalk. (2015). Firmware Analysis Tool. Retrieved on December 8, 2015, from http://binwalk.org/

[6] Blunden, B. (2009). The Rootkit Arsenal: Escape and Evasion in the Dark Corners of the System. 1st edition. Jones & Bartlett Learning.

[7] Blunden, B. (2012). The Rootkit Arsenal: Escape and Evasion in the Dark Corners of the System. 2nd edition. Burlington, MA: Jones & Bartlett Publishers.

[8] Brown, W. (2009). Building and Using an Automated Malware Analysis Pipeline Tools, Techniques, and Mindset. Proceedings of the Hack in The Box Security Conference (HITB), Malaysia. Retrieved on December 8, 2015, from http://conference.hitb.org/hitbsecconf2009kl/materials/D2T3%20-%20Wes%20Brown%20-%20Building%20and%20Using%20an%20Automated%20Malware%20Analysis%20Pipeline.pdf

[9] Calvet, J. (2012). Cryptographic Function Identification in Obfuscated Binary Programs. Proceedings of the REcon. Montreal, Canada. Retrieved on December 8, 2015, from http://recon.cx/2012/schedule/attachments/46_Joan_CryptographicFunctionIdentification.pdf

[10] Chen, L. (2008). Zion system DKOM driver kernel. Microsoft Corporation. Retrieved on December 8, 2015, from http://blogs.technet.com/cfs-filesystemfile.ashx/__key/telligent-evolution-components-attachments/01-6336-00-00-03-07-17-55/Zion.zip

[11] Chen, R. (2004). The history of calling conventions, part 3. MSDN. Retrieved on December 8, 2015, from https://blogs.msdn.microsoft.com/oldnewthing/20040108-00/?p=41163/

[12] Choi, Y., Kim, I., Oh, J., & Ryou, J. (2009). Encoded Executable File Detection Technique via Executable File Header Analysis. International Journal of Hybrid Information Technology. 2(2). Retrieved on December 8, 2015, from http://www.sersc.org/journals/IJHIT/vol2_no2_2009/3.pdf

[13] Ch40zz. (2015). PspCidTable and Patchguard on x64. Rohitab. Retrieved from http://www.rohitab.com/discuss/topic/41909-pspcidtable-and-patchguard-on-x64/?p=10101659

[14] Clark, B. (2014). Unlinking Windows Services from the Service Record List. Retrieved on December 8, 2015, from http://speakingofcyber.blogspot.ru/2014/08/unlinking-windows-services-from-service.html

[15] Cohen, M. (2014).Pooltags for common objects. Rekall Memory Forensics Retrieved on December 8, 2015, from http://www.rekall-forensic.com/epydocs/rekall-module.html

[16] Conti, G., Bratus, S., Shubina, A., Sangster, B., Ragsdale, R., Supan, M., Lichtenberg, A.. & Perez-Alemany, R. (2010). Automated Mapping of Large Binary Objects Using Primitive Fragment Type Classification. The International Journal of Digital Forensics & Incident Response. Volume 7. Pages S3-S12. Elsevier Science Publishers. Amsterdam, The Netherlands.

[17] Cortesi, A. (2012) Visualizing Entropy in Binary Files. Retrieved on December 8, 2015, from http://corte.si/posts/visualisation/entropy/index.html

[18] Cortesi, A. (2015a). A Browser-based Tool for Visualizing Binary Data. Retrieved on December 8, 2015, from http://binvis.io/#/

[19] Cortesi, A. (2015b). BinVis tool download. Retrieved on December 8, 2015, from http://www.rumint.org/gregconti/publications/binviz_0.zip

[20] Dang, B., Gazet, A., Bachaalany, E., & Josse, S. (2014). Practical Reverse Engineering: x86, x64, ARM, Windows Kernel, Reversing Tools, and Obfuscation. Indianapolis, IN, USA: John Wiley & Sons, 1 edition, 384 p.

[21] Devi, D., & Nandi, S. (2012). PE File Features in Detection of Packed Executables. International Journal of Computer Theory and Engineering. (4)3. Retrieved on December 8, 2015, from http://www.ijcte.org/papers/512-S10014.pdf

[22] Ding, Y., Dai, W., Yan, S., & Zhang, Y. (2014). Control Flow-based Opcode Behavior Analysis for Malware Detection. Computers & Security. Volume 44. 65–74. http://dx.doi.org/10.1016/j.cose.2014.04.003

[23] Dingbaum, S. (2016). Audit Of NRC'S Network Security Operations Center. United States Nuclear Regulatory Commission. Reference # OIG-16-A-07. Retrieved from http://pbadupws.nrc.gov/docs/ML1601/ML16011A319.pdf

[24] Dorfman, A. (2014). FRODO: Format Reverser of Data Objects. Proceedings of the Hack In The Box Security Conference (HITB), Amsterdam, Netherlands. Retrieved on December 8, 2015, from http://bofh.nikhef.nl/events/HitB/hitb-2014-amsterdam/praatjes/D2T3-Format-Reverser-of-Data-Objects.pdf

[25] Eagle, C. (2011). The IDA Pro Book: The Unofficial Guide to the World's Most Popular Disassembler. No Starch Press. 2nd Edition. 672p.

[26] Emil's Projects. (2010). Tiny C Scrambling Compiler. OpenHardware & OpenSource. Retrieved on December 8, 2015, from http://uglyduck.ath.cx/ep/archive/2010/04/Tiny_C_Scrambling_Compiler.html

[27] Farrukh, S., & Muddassar, F. (2012). ELF-Miner: Using Structural Knowledge and Data Mining Methods To Detect New (Linux) Malicious Executables. Knowledge and Information Systems. 30 (3), 589-612 Springer-Verlag New York, NY, USA http://dx.doi.org/10.1007/s10115-011-0393-5

[28] Gilbert, B. (2015). ISIS Planning Major Cyberattacks Against Airlines, Hospitals And Nuclear Power Plants. International Business Times. Retrieved from http://www.ibtimes.com/isis-planning-major-cyberattacks-against-airlines-hospitals-nuclear-power-plants-2187567

[29] Ghezelbigloo, Z., & VafaeiJahan, M. (2014). Role-opcode vs. Opcode: the New method in Computer Malware Detection. International Congress on Technology, Communication and Knowledge (ICTCK). 1-6. Mashhad. Iran. http://dx.doi.org/10.1109/ICTCK.2014.7033534

[30] Hall, G.A., & Davis, W.P. (2006). Sliding Window Measurement for File Type Identification. Technical report, Computer Forensics and Intrusion Analysis Group, ManTech. Security and Mission Assurance, Rexas

[31] Haukli, L. (2014). Exposing Bootkits with BIOS Emulation. Black Hat USA. Retrieved on December 8, 2015, from https://www.blackhat.com/docs/us-14/materials/us-14-Haukli-Exposing-Bootkits-With-BIOS-Emulation.pdf

[32] hfiref0x. (2016). Driver loader for bypassing Windows x64 Driver Signature Enforcement. GitHub. Retrieved from https://github.com/hfiref0x/TDL

[33] Hoglund, G., & Butler, J. (2005). Direct Kernel Object Manipulation. Rootkits: Subverting the Windows Kernel. 169-212. Addison-Wesley Professional.

[34] Iwamoto, K., & Wasaki, K. (2014). A Method for Shellcode Extraction from Malicious Document Files Using Entropy and Emulation. IACSIT International Journal of Engineering and Technology. (8)2. 101-106. Singapore, http://dx.doi.org/10.7763/IJET.2016.V8.866

[35] Jakobsson, M., & Ramzan, Z. (2008). Direct Kernel Object Manipulation. Crimeware: Understanding New Attacks and Defenses. 1st edition. 253-254. Addison-Wesley Professional.

[36] Jason, A. (2012). Ghost in the Shell: A Counter-intelligence Method for Spying while Hiding in (or from) the Kernel with APCs. Thesis. Queen's University. Kingston, Ontario, Canada. Retrieved on December 8, 2015, from https://qspace.library.queensu.ca/bitstream/1974/7605/1/Alexander_Jason_S_201210_MSC.pdf

[37] Jeff, J. (2015). Cyberattack on US Power Plants Could Cause $1T in Economic Damage. Green Technology. Retrieved

from http://www.greentechmedia.com/articles/read/cyber-attack-on-u.s.-power-plants-could-cause-1t-in-economic-damage

[38] Jochheim, B. (2012). On the Automatic Detection of Embedded Malicious Binary Code using Signal Processing Techniques. Project Report. Retrieved on December 8, 2015, from http://inet.cpt.haw-hamburg.de/teaching/ss-2012/master-projects/benjamin_jochheim_pr1.pdf

[39] Jochheim, B. (2012, October 17). On the Automatic Detection of Embedded Malicious Binary Code using Signal Processing Techniques. Project Report. Retrieved on December 8, 2015, from https://inet.cpt.haw-hamburg.de/teaching/ss-2012/master-projects/benjamin_jochheim_pr1.pdf

[40] Jochheim, B., Schmidt, T.C., & Wahlisch, M. (2011). A Signature-free Approach to Malicious Code Detection by Applying Entropy Analysis to Network Streams. Project SKIMS. Retrieved on December 8, 2015, from https://tnc2011.terena.org/getfile/489

[41] Kaspersky Lab. (2015, June 11). The Duqu 2.0 Technical Details. Retrieved on December 8, 2015, from https://securelist.com/files/2015/06/The_Mystery_of_Duqu_2_0_a_sophisticated_cyberespionage_actor_returns.pdf

[42] Kaspersky Lab's Global Research & Analysis Team. (2013, January 14). “Red October” Diplomatic Cyber Attacks Investigation. Retrieved on December 8, 2015, from https://securelist.com/analysis/publications/36740/red-october-diplomatic-cyber-attacks-investigation/

[43] Kaspersky, E. (2015, June 10). Kaspersky Lab investigates hacker attack on its own network. Retrieved on December 8, 2015, from https://blog.kaspersky.com/kaspersky-statement-duqu-attack/8997/

[44] KernelMode.info. (2012). [Kernel] Unloaded modules list. Forum discussion. Retrieved on December 8, 2015, from http://www.kernelmode.info/forum/viewtopic.php?t=1549

[45] Kohli, P., Lempitsky, V., & Barinova. O. (2015). Detecting and Localizing Multiple Objects in Images Using Probabilistic Inference, U.S. Patent No. US8953888 B2. Washington, DC: U.S. Patent and Trademark Office.

[46] Korkin, I., & Nesterov I. (2014, May 28-29). Applying Memory Forensics to Rootkit Detection. Paper presented at the Proceedings of the 9th annual Conference on Digital Forensics, Security and Law (CDFSL), 115-141, Richmond, VA, USA.

[47] Korkin, I. (2012). Windows 8 is Cyber-Battlefield. Retrieved on December 8, 2015, from http://www.igorkorkin.blogspot.ru/2012/09/windows-8-is-cyber-battlefield.html

[48] Ladakis, E., Koromilas, L., Vasiliadis, G., Polychronakis, M., & Ioannidis, S. (2013).You Can Type, but You Can’t Hide: A Stealthy GPU-based Keylogger. Proceedings of the 6th European Workshop on System Security (EuroSec). Prague, Czech Republic. Retrieved on December 8, 2015, from http://www.cs.columbia.edu/~mikepo/papers/gpukeylogger.eurosec13.pdf

[49] Ligh MH., Case, A., Levy, J., & Walters, A. (2014a). Detecting Orphan Threads. The Art of Memory Forensics: Detecting malware and threats in Windows, Linux, and Mac memory. 379-380. 1st edition. Wiley. Indianapolis, IN, USA.

[50] Ligh MH., Case, A., Levy, J., & Walters, A. (2014b). Recently Unloaded Modules. The Art of Memory Forensics: Detecting malware and threats in Windows, Linux, and Mac Memory. 374-375. 1st edition. Wiley. Indianapolis, IN, USA.

[51] Ligh MH., Case, A., Levy, J., & Walters, A. (2014c). Windows Services. The Art of Memory Forensics: Detecting malware and threats in Windows, Linux, and Mac Memory. 343-366. 1st edition. Wiley. Indianapolis, IN, USA.

[52] Ligh MH., Case, A., Levy, J., & Walters, A. (2014d). Revealing Hidden Services. The Art of Memory Forensics: Detecting malware and threats in Windows, Linux, and Mac Memory. 362-366. 1st edition. Wiley. Indianapolis, IN, USA.

[53] Ligh MH., Case, A., Levy, J., & Walters, A. (2014e). Revealing Hidden Services. The Art of Memory Forensics: Detecting malware and threats in Windows, Linux, and Mac memory.

362-366. 1st edition. Wiley. Indianapolis, IN, USA.

[54] Ligh MH., Case, A., Levy, J., & Walters, A. (2014f). Scanning Memory. The Art of Memory Forensics: Detecting malware and threats in Windows, Linux, and Mac Memory. 351-352. 1st edition. Wiley. Indianapolis, IN, USA.

[55] Ligh MH., Case, A., Levy, J., & Walters, A. (2014g). Volatility's SvcScan Plugin. The Art of Memory Forensics: Detecting malware and threats in Windows, Linux, and Mac Memory. 352-353. 1st edition. Wiley. Indianapolis, IN, USA.

[56] Ligh MH., Case, A., Levy, J., & Walters, A. (2014h). Revealing Hidden Services. The Art of Memory Forensics: Detecting malware and threats in Windows, Linux, and Mac Memory. 362-366. 1st edition. Wiley. Indianapolis, IN, USA.

[57] Ligh MH., Case, A., Levy, J., & Walters, A. (2014i). Stealthy Hooks. The Art of Memory Forensics: Detecting malware and threats in Windows, Linux, and Mac Memory. 384-386. 1st edition. Wiley. Indianapolis, IN, USA.

[58] Ligh, MH. (2015a). PlugX: The Memory Forensics Lifecycle. Retrieved on December 8, 2015, from https://prezi.com/6ruvzpnpp-8y/plugx-the-memory-forensics-lifecycle/

[59] Ligh, MH.. (2015b). ScInitDatabase signature. Retrieved on December 8, 2015, from https://github.com/volatilityfoundation/volatility/blob/master/volatility/plugins/malware/servicediff.py

[60] Ligh, MH.., Adair, S., Hartstein, B., & Richard, M. (2010). The Case of Blazgel. Malware Analyst's Cookbook and DVD: Tools and Techniques for Fighting Malicious Code. 664-669. 1st edition. Wiley.

[61] Ligh, MH. (2011). Investigating Windows Threads with Volatility. MNIN Security Blog. Retrieved on December 8, 2015, from http://mnin.blogspot.ru/2011/04/investigating-windows-threads-with.html

[62] Ligh, MH., Case, A., Levy, J., & Walters, A. (2014, July 28). The Art of Memory Forensics: Detecting Malware and Threats in Windows, Linux, and Mac Memory. 1st edition. 912 pp. Wiley Publishing.

[63] Linchpin Labs (2010). ATSIV utility. Retrieved on December 8, 2015, from http://www.linchpinlabs.com

[64] Louboutin, C. (2010). Service Hiding. Retrieved on December 8, 2015, from http://club.1688.com/article/12193937.htm

[65] Lupi, V. (2011). How to Write a Good Rootkit: a Different Approach. Hakin9 Extra Magazine, English Edition. 06. 18-21. Retrieved on December 8, 2015, from http://codelaughs.blogspot.it/2011/11/how-to-write-good-rootkit-different.html

[66] Lyda, R., & Hamrock, J. (2007). Using Entropy Analysis to Find Encrypted and Packed Malware. Journal IEEE Security and Privacy. 5(2). 40-45. Piscataway, NJ, USA. http://dx.doi.org/10.1109/MSP.2007.48

[67] Maartmann-Moe, C. (2008). Forensic Key Discovery and Identification. Master of Science in Communication Technology. Norwegian University of Science and Technology. Retrieved on December 8, 2015, from http://www.diva-portal.org/smash/get/diva2:347635/FULLTEXT01.pdf

[68] Marcos, M. (2014). Without a Trace: Fileless Malware Spotted in the Wild. Trend Micro. Security Intelligence Blog. Retrieved on December 8, 2015, from http://blog.trendmicro.com/trendlabs-security-intelligence/without-a-trace-fileless-malware-spotted-in-the-wild/

[69] Mask, A. (2011). Service Hiding. Retrieved on December 8, 2015, from http://maskattached.blogspot.ru/2011/04/service-hiding.html

[70] Matrosov, A., Rodionov, E., & Bratus, S. (2016). Rootkits and Bootkits. Reversing Modern Malware and Next Generation Threats. ISBN: 978-1-59327-716-1. 304 pp. No Starch Press.

[71] Matveeva, V., & Epishkina, A. (2015). Searching for Random Data in File System During Forensic Expertise. Biosciences, Biotechnology Research Asia, BBRA. 12(1). 745-752. http://dx.doi.org/10.13005/bbra/1720

[72] Matveeva, V. (2014). Information Entropy and Its Application for Information Security Tasks. Security of Information Technologies, Issue #4, ISSN 2074-7128, 30-36.

[73] McAfee Labs. (2015a). Threats Report. Retrieved on December 8, 2015, from http://www.mcafee.com/us/resources/reports/rp-quarterly-threat-q1-2015.pdf

[74] McAfee Labs. (2015b). 2016 Threats Predictions report. Retrieved on December 8, 2015, from http://www.mcafee.com/us/resources/reports/rp-threats-predictions-2016.pdf

[75] Merkel, R., Hoppe, T., Kraetzer, C., & Dittmann, J. (2010). Statistical Detection of Malicious PE-Executables for Fast Offline Analysis. Proceedings of Communications and Multimedia Security. 93-105. Linz. Austria. http://dx.doi.org/10.1007/978-3-642-13241-4_10

[76] Microsoft. (n.d.-a). File sysload.c. Windows Research Kernel Source Code. Retrieved on December 8, 2015, from http://gate.upm.ro/os/LABs/Windows_OS_Internals_Curriculum_Resource_Kit-ACADEMIC/WindowsResearchKernel-WRK/WRK-v1.2/base/ntos/mm/sysload.c

[77] Microsoft. (n.d.-b). File mm.h. Windows Research Kernel Source Code. Retrieved on December 8, 2015, from http://gate.upm.ro/os/LABs/Windows_OS_Internals_Curriculum_Resource_Kit-ACADEMIC/WindowsResearchKernel-WRK/WRK-v1.2/base/ntos/inc/mm.h

[78] Microsoft. (n.d.-c). CurrentControlSet\Services Subkey Entries. https://support.microsoft.com/en-us/kb/103000

[79] Microsoft. (n.d.-d). File windows_nt_4_source_code_IK\nt4\private\windows\screg\sc\server\dataman.c. Windows NT 4.0 Full Free Source Code. Retrieved on December 8, 2015, from http://igorkorkin.blogspot.ru/2013/09/windows-nt-40-full-free-source-code-912_16.html

[80] Microsoft. (n.d.-e). File windows_nt_4_source_code_IK\nt4\private\ntos\dd\kbdclass\kbdclass.h. Windows NT 4.0 Full Free Source Code. Retrieved on December 8, 2015, from http://igorkorkin.blogspot.ru/2013/09/windows-nt-40-full-free-source-code-912_16.html

[81] Microsoft. (n.d.-f). File windows_nt_4_source_code_IK\nt4\private\ntos\dd\i8042prt\i8042prt.h. Windows NT 4.0 Full Free Source Code. Retrieved on December 8, 2015, from http://igorkorkin.blogspot.ru/2013/09/windows-nt-40-full-free-source-code-912_16.html

[82] MidnightCowboy (2015, May 25). Best Free Rootkit Scanner and Remover. Retrieved on December 8, 2015, from http://www.techsupportalert.com/best-free-rootkit-scanner-remover.htm

[83] Milkovic, L. (2012). DriverHider. Dementia Project. Retrieved on December 8, 2015, from http://dementia-forensics.googlecode.com/svn/trunk/DementiaKM/DriverHider.cpp

[84] MJ0011. (2009). Analysis OS and Detection Rootkit Outside the VMWare. Retrieved on December 8, 2015, from http://powerofcommunity.net/poc2009/mj.pdf

[85] MSDN. (n.d.-a). NtQueryDirectoryObject function. Retrieved on December 8, 2015, from https://msdn.microsoft.com/en-us/library/bb470238(v=vs.85).aspx

[86] MSDN. (n.d.-b). IDebugDataSpaces::ReadDebuggerData method. Retrieved on December 8, 2015, from https://msdn.microsoft.com/en-us/library/windows/hardware/ff553536(v=vs.85).aspx

[87] MSDN. (n.d.-c). EnumServicesStatus function. Retrieved on December 8, 2015, from https://msdn.microsoft.com/en-us/library/windows/desktop/ms682637(v=vs.85).aspx

[88] MSDN. (n.d.-d). Introduction to Registry Keys for Drivers. Retrieved on December 8, 2015, from https://msdn.microsoft.com/en-us/library/windows/hardware/ff544262(v=vs.85).aspx

[89] Musavi, S.A., & Kharrazi, M. (2014). Back to Static Analysis for Kernel-Level Rootkit Detection. IEEE Transactions on Information Forensics and Security. (9)9. 1465-1476. http://dx.doi.org/10.1109/TIFS.2014.2337256

[90] Nataraja, L., Jacobb, G., & Manjunatha, B.S. (2010). Detecting Packed Executables based on Raw Binary Data. Technical Report. Retrieved on December 8, 2015, from https://vision.ece.ucsb.edu/sites/vision.ece.ucsb.edu/files/publications/packed-unpacked-tech-report.pdf

[91] Nath, H.V., & Mehtre, B.M. (2014, March 13-14). Static Malware Analysis Using

Machine Learning Methods. Recent Trends in Computer Networks and Distributed Systems Security. Proceedings of the Second International Conference, SNDS 2014, Trivandrum, India. 440-450. http://dx.doi.org/10.1007/978-3-642-54525-2_39

[92] Nguyen, A. (2014, August 7). Capstone: Next-Gen Disassembly Framework. Blackhat USA. Retrieved on December 8, 2015, from http://capstone-engine.org/BHUSA2014-capstone.pdf

[93] Nguyen, K., Tran, D., Ma, W., & Sharma, D. (August 19-21, 2014). An Approach to Detect Network Attacks Applied for Network Forensics. 11th International Conference on Fuzzy Systems and Knowledge Discovery (FSKD). 655-660. Xiamen, China. http://dx.doi.org/10.1109/FSKD.2014.6980912

[94] NTInfo. (2014). Entropy and the distinctive signs of packed PE files. Retrieved on December 8, 2015, from http://n10info.blogspot.ru/2014/06/entropy-and-distinctive-signs-of-packed.html

[95] OSR Online. (1997). Using the NT Registry for Driver Install. Retrieved on December 8, 2015, from https://www.osronline.com/article.cfm?id=170

[96] Paganini, P. (2015, August 28). Symantec discovered 49 New Modules of the Regin espionage platform. Retrieved on December 8, 2015, from http://securityaffairs.co/wordpress/39647/cyber-crime/symantec-49-new-regin-modules.html

[97] Paganini, P. (June 17, 2015). Duqu 2.0: The Most Sophisticated Malware Ever Seen. Retrieved on December 8, 2015, from http://resources.infosecinstitute.com/duqu-2-0-the-most-sophisticated-malware-ever-seen

[98] Pareek, H., Eswari, P. R. L., & Babu S.C. (2013). Entropy and n-gram Analysis of Malicious PDF Documents. International Journal of Engineering. 2(2). Retrieved on December 8, 2015, from https://www.researchgate.net/publication/235974671_Entropy_and_n-gram_analysis_of_malicious_PDF_documents

[99] Pedrero, X.-U., Santos, I., Sanz. B., Laorden, C., & Bringas P.-G. (2012, January 14-17). Countering Entropy Measure Attacks on Packed Software Detection. Proceedings of the 9th IEEE Consumer Communications and Networking Conference (CCNC2012), Las Vegas, NV, USA. http://dx.doi.org/10.1109/CCNC.2012.6181079

[100] Pistelli, D. (n.d.). AntiMida 1.0. Retrieved on December 8, 2015, from http://www.ntcore.com/files/antimida_1.0.htm

[101] Probert, D. (2004). Windows Kernel Internals Object Manager & LPC. Retrieved on December 8, 2015, from http://i-web.i.u-tokyo.ac.jp/edu/training/ss/msprojects/data/04-ObjectManagerLPC.ppt

[102] Qark. (n.d.). Windows Executable Infection. VX Heaven. Retrieved on December 8, 2015, from http://vxheaven.org/lib/static/vdat/tuvd0004.htm

[103] Ragan, S. (2016). Israel's electric grid targeted by malware, energy minister says. CSO Online. Retrieved from http://www.csoonline.com/article/3026604/security/israels-electric-grid-targeted-by-malware-energy-minister-says.html

[104] Reactor pattern. (n.d.). In Wikipedia. Retrieved on December 8, 2015, from https://en.wikipedia.org/wiki/Reactor_pattern

[105] ReactOS. (n.d.). KBDHID_DEVICE_EXTENSION Struct Reference. Retrieved on December 8, 2015, from http://doxygen.reactos.org/da/daa/structKBDHID__DEVICE__EXTENSION.html

[106] Rhee, Y. (2009). Pool tag list. TechNet Blogs. Retrieved on December 8, 2015, from http://blogs.technet.com/b/yongrhee/archive/2009/06/24/pool-tag-list.aspx

[107] Russinovich, M., Solomon, D., & Ionescu, A. (2012). Windows Internals, Parts 1 and 2 (Developer Reference). 6th edition. Microsoft Press.

[108] Russinovich, M. (2011). Analyzing a Stuxnet Infection with the Sysinternals Tools, Part 1. Microsoft TechNet. Retrieved on December 8, 2015, from https://blogs.technet.microsoft.com/markrussinovich/2011/03/26/analyzing-a-stuxnet-infection-with-the-sysinternals-tools-part-1/

[109] Saleh, M., Ratazzi E.P., & Xu S, (2014). Instructions-based Detection of Sophisticated Obfuscation and Packing. Proceedings of the

33rd Annual IEEE Military Communications Conference, MILCOM, 1-6, Baltimore, MD, USA. dx.doi.org/10.1109/MILCOM.2014.9

[110] SANS Institute (2014). A Case Study of an Incident: A Day in the Life of an IR Team. Retrieved on December 8, 2015, from http://computerforensicsblog.champlain.edu/wp-content/uploads/2014/06/APT-Attacks-Exposed-Network-Host-Memory-and-Malware-Analysis-Lee-Tilbury-Hagen-5-21-2014.pdf

[111] SANS Institute (2015b). How to Parse a Memory Image with the Volatility Framework. Sans Memory Forensics Poster. Retrieved on December 8, 2015, from http://digital-forensics.sans.org/media/Poster-2015-Memory-Forensics2.pdf

[112] SANS Institute. (2015a). Rise of anti-forensics techniques requires response from digital investigators [Press release]. Retrieved on December 8, 2015, from https://www.sans.org/press/announcement/2015/09/07/1

[113] Santos, I., Brezo, F., Nieves, J. K., Penya, Y.K., Sanz, B., Laorden, C., & Bringas, P.G. (2010). Idea: Opcode-sequence-based Malware Detection. In Engineering Secure Software and Systems, Second International Symposium, ESSoS, F. Massacci, D. S. Wallach, and N. Zannone, Eds., vol. 5965 of Lecture Notes in Computer Science, Springer, 35-43. http://dx.doi.org/10.1007/978-3-642-11747-3_3. Retrieved on December 8, 2015, from http://paginaspersonales.deusto.es/ypenya/publi/penya_ESSoS2010_Opcode-sequence-based%20Malware%20Detection.pdf

[114] Santos, I., Brezo, F., Ugarte-Pedrero, X., & Bringas, P.G. (2013). Opcode Sequences as Representation of Executables for Data-mining-based Unknown Malware Detection. Information Sciences. 231. 64-82. Elsevier Science. New York, NY, USA. http://dx.doi.org/10.1016/j.ins.2011.08.020

[115] Santos, I., Sanz, B., Laorden, C., Brezo, F., & Bringas, P.G. (2011). Opcode-sequence-based semi-supervised unknown malware detection. Proceedings of the 4th International Conference on Computational Intelligence in Security for Information Systems, CISIS'11. 50-57. Berlin, Heidelberg. http://dx.doi.org/10.1007/978-3-642-21323-6_7

[116] Santos, R. (2014). Poweliks: Malware Hides in Windows Registry. Trend Micro. Security Intelligence Blog. Retrieved on December 8, 2015, from http://blog.trendmicro.com/trendlabs-security-intelligence/poweliks-malware-hides-in-windows-registry/

[117] Schmidt, T., Wahlisch, M., & Groning, M. (2011, June 15-17). Context-adaptive Entropy Analysis as a Lightweight Detector of Zero-day Shellcode Intrusion for Mobiles. Poster at The ACM Conference on Wireless Network Security (WiSec). Hamburg, Germany.

[118] Schuberth, T. (2014). Modern Threats and Malware and IT-Security's Future. Retrieved on December 8, 2015, from http://www.swisst.net/files/swisstnet/de/dokumente/Communication_Conference_2014/aktuelle_gefahren_und_die_zukunft_der_it_security.pdf

[119] Schuster, A. (2006). Searching for processes and threads in Microsoft Windows memory dumps. Journal Digital Investigation: The International Journal of Digital Forensics & Incident Response. 3, 10-16. http://dx.doi.org/10.1016/j.diin.2006.06.010

[120] Seals, T. (2015, September 8). Anti-Forensic Malware Widens Cyber-Skills Gap. Retrieved on December 8, 2015, from http://www.infosecurity-magazine.com/news/antiforensic-malware-widens

[121] Shabtai, A., Moskovitch, R., Feher, C., Dolev, S., & Elovici, Y. (2012). Detecting Unknown Malicious Code by Applying Classification Techniques on OpCode Patterns. Security Informatics. 1(1). Retrieved on December 8, 2015, from http://www.security-informatics.com/content/pdf/2190-8532-1-1.pdf

[122] Siddiqui, M., Wang, M.C., & Lee., J. (2008). Data Mining Methods for Malware Detection Using Instruction Sequences. Proceedings of the 26th IASTED International Conference on Artificial Intelligence and Applications, (AIA'08). 358-363. Anaheim, CA, USA. ACTA Press.

[123] Silcott, G., & Swinimer, J. (2015). AMD Launches 'Boltzmann Initiative' to

Dramatically Reduce Barriers to GPU Computing on AMD FirePro Graphics. [Press Releases]. Retrieved on December 8, 2015, from http://www.amd.com/en-us/press-releases/Pages/boltzmann-initiative-2015nov16.aspx

[124] Singla, S., Gandotra, E., Bansal, D., & Sofat, S. (2015). A Novel Approach to Malware Detection using Static Classification. International Journal of Computer Science and Information Security (IJCSIS). 13(3). Retrieved on December 8, 2015, from https://www.academia.edu/11754857/A_Novel_Approach_to_Malware_Detection_using_Static_Classification

[125] Stewin, P., & Bystrov, I. (2012). Understanding DMA Malware. Paper presented at Proceedings of the 9th Conference on Detection of Intrusions and Malware & Vulnerability Assessment. Heraklion, Crete, Greece. Retrieved on December 8, 2015, from http://www.stewin.org/papers/dimvap15-stewin.pdf

[126] Stuttard, D., Pinto, M., Ligh, M.H., Adair, S., Hartstein, B., & Richard, M. (2014, January 28). Attack and Defend Computer Security Set. 1st Edition. Wiley.

[127] Suszter, A. (2014). Examining Unknown Binary Formats Retrieved on December 8, 2015, from http://reversingonwindows.blogspot.ru/2014/04/examining-unknown-binary-formats.html

[128] Symantec. (2015, August 27). Regin: Top-tier espionage tool enables stealthy surveillance. Retrieved on December 8, 2015, from http://www.symantec.com/content/en/us/enterprise/media/security_response/whitepapers/regin-analysis.pdf

[129] Tabish, S.M., Shafiq, M.Z., & Farooq, M. (2009). Malware Detection using Statistical Analysis of Byte-Level File Content. Proceedings of the ACM SIGKDD Workshop on CyberSecurity and Intelligence Informatics (CSI-KDD'09). 23-31. http://dx.doi.org/10.1145/1599272.1599278 New York, NY, USA

[130] Tanda, S. (2015). GitHub. Retrieved on December 8, 2015, from https://github.com/tandasat

[131] Teller, T. (2013). Detecting the One Percent: Advanced Targeted Malware Detection. Proceedings of the RSA Conference. San Francisco. USA. Retrieved on December 8, 2015, from https://www.rsaconference.com/writable/presentations/file_upload/spo2-t19_spo2-t19.pdf

[132] The Register. (2015, October 8). New mystery Windows-smashing RAT found in corporate network. Retrieved on December 8, 2015, from http://www.theregister.co.uk/2015/10/08/monker_rat/

[133] Timzen, T. (2015). Kernel Forensics and Rootkits. Course material for Computer Forensics III (Memory Forensics) / CS407. [Lecture notes]. Retrieved on December 8, 2015, from https://www.tophertimzen.com/resources/cs407/slides/week06_01-Rootkits.html#slide1

[134] Tsaur, W.J., & Chen, Y.C. (2010). Exploring Rootkit Detectors' Vulnerabilities Using a New Windows Hidden Driver Based Rootkit. Paper presented at The Second IEEE International Conference on Social Computing (SocialCom2010), Minneapolis, MN, USA. 842-848. http://dx.doi.org/10.1109/SocialCom.2010.127

[135] Tsaur, W.J., & Wu, J.X. (2014b). Removing the driver-related information in the registry. New Windows Rootkit Technologies for Enhancing Digital Rights Management in Cloud Computing Environments. Proceedings of the 2014 International Conference on e-Learning, e-Business, Enterprise Information Systems, and e-Government (EEE'14). Las Vegas, USA. 3-4. Retrieved on December 8, 2015, from http://worldcomp-proceedings.com/proc/p2014/EEE2315.pdf

[136] Tsaur, W.J., & Wu, J.X. (2014c). Removing the Signature of PE (Portable Executable) Image. New Windows Rootkit Technologies for Enhancing Digital Rights Management in Cloud Computing Environments. 2-3. Proceedings of the 2014 International Conference on e-Learning, e-Business, Enterprise Information Systems, and e-Government (EEE'14). Las Vegas, USA. Retrieved on December 8, 2015, from http://worldcomp-proceedings.com/proc/p2014/EEE2315.pdf

[137] Tsaur, W.J., & Yeh, L.Y. (2015, July 27-30). New Protection of Kernel-level Digital Rights Management in Cloud-based Consumer Electronics Environments. Proceedings of the 2015 International Conference on Grid & Cloud Computing and Applications (GCA'15). Monte Carlo Resort, Las Vegas, USA. Retrieved on December 8, 2015, from http://worldcomp-proceedings.com/proc/p2015/GCA2880.pdf

[138] Tsaur, W.J., & Wu, J.X. (2014a). Removing Drivers from PsLoadedModuleList. New Windows Rootkit Technologies for Enhancing Digital Rights Management in Cloud Computing Environments. Proceedings of the 2014 International Conference on e-Learning, e-Business, Enterprise Information Systems, and e-Government (EEE'14). Las Vegas, USA. Retrieved on December 8, 2015, from http://worldcomp-proceedings.com/proc/p2014/EEE2315.pdf

[139] Visualizing ELF binaries. (2014). Reverse Engineering. Retrieved on December 8, 2015, from http://reverseengineering.stackexchange.com/questions/6003/visualizing-elf-binaries

[140] Vomel, S., & Lenz, H. (2013, March 12-14). Visualizing Indicators of Rootkit Infections in Memory Forensics, Paper presented at 7th International Conference on IT Security Incident Management and IT Forensics (IMF), 122-139, Nuremberg, German.

[141] Wang, T.Y., Wu, C.H., & Hsieh, C.C. (2009). Detecting Unknown Malicious Executables Using Portable Executable Headers. Proceedings of the 2009 Fifth International Joint Conference on INC, IMS and IDC. 278-284. Washington, DC, USA. http://dx.doi.org/10.1109/NCM.2009.385

[142] Wangen, G. (2015, May 18). The Role of Malware in Reported Cyber Espionage: A Review of the Impact and Mechanism. Information. 6(2). 183-211. http://dx.doi.org/10.3390/info6020183

[143] Weil, N. (2014, November 24). Stealthy, sophisticated 'Regin' malware has been infecting computers since 2008. Retrieved on December 8, 2015, from http://www.pcworld.com/article/2851472/symantec-identifies-sophisticated-stealthy-regin-malware.html

[144] Wineblat, E. (2009). Service Hiding. Apriorit Inc. Retrieved on December 8, 2015, from http://www.codeproject.com/Articles/46670/Service-Hiding

[145] x86 Disassembly/Windows Executable Files. (n.d.). In Wikipedia. Retrieved on December 8, 2015, from https://en.wikibooks.org/wiki/X86_Disassembly/Windows_Executable_Files

[146] Yu, S., Zhou, S., Liu, L., Yang, R., & Luo, J. (2011). Detecting Malware Variants by Byte Frequency. 2010 Second International Conference on Networks Security Wireless Communications and Trusted Computing (NSWCTC). 32-35. Wuhan, Hubei, China. http://dx.doi.org/10.1109/NSWCTC.2010.145

[147] Yurichev, D. (2015). Analyzing unknown binary files using information entropy. Retrieved on December 8, 2015, from http://yurichev.com/blog/entropy/

[148] ZCM Services. (2010). Manually Adding or Removing Services and Devices. Retrieved on December 8, 2015, from http://nt4ref.zcm.com.au/mansd.htm

[149] zhanglinfu2000. (2013). The organization of the Windows object. Retrieved on December 8, 2015, from http://www.developermemo.com/3524027/

[150] Zolotukhin, M. & Hamalainen, T. (2014). Detection of Zero-Day Malware Based On the Analysis of Opcode Sequences. Proceedings of the 11th Consumer Communications and Networking Conference (CCNC). 386-391. Las Vegas, NV, USA. http://dx.doi.org/10.1109/CCNC.2014.6866599